\documentclass[preprint,11pt]{elsarticle}

\usepackage{graphicx}
\usepackage{amssymb}
\usepackage{epstopdf}
\usepackage{bm}
\usepackage{color}

\usepackage[latin1]{inputenc}
\usepackage[T1]{fontenc}
\usepackage[english]{babel}
\usepackage[dvipsnames]{xcolor}
\usepackage{color}
\usepackage{geometry}
\usepackage{capt-of}
\usepackage{pstricks}
\usepackage{bm}
\usepackage{slashed}
\usepackage{booktabs}

\usepackage{bm,amsmath,amssymb}

\long\def\comment#1{ }

\newcommand{\beq}{\begin{eqnarray}}
\newcommand{\eeq}{\end{eqnarray}}
\newcommand{\nn}{\nonumber}

\newcommand{\rmd}{{\rm d}}
\newcommand{\rme}{{\rm e}}

\def\q{{\bm q}}
\def\p{{\bm p}}

\def\A{{\boldsymbol A}}

\def\bmg{{\boldsymbol g}}
\def\bmh{{\boldsymbol h}}

\def\k{{\boldsymbol k}}

\def\x{{\boldsymbol x}}
\def\y{{\boldsymbol y}}

\def\D{{\boldsymbol D}}
\def\r{{\boldsymbol r}}
\def\z{{\boldsymbol z}}

\def\v{{\boldsymbol v}}

\def\Q{{\boldsymbol Q}}

\def\bmrho{{\boldsymbol \rho}}

\def\bmgamma{\boldsymbol \gamma}

\def\nab{{\boldsymbol \nabla}}
\def\balpha{{\boldsymbol \alpha}}

\newcommand{\bms}{\bm{s}}

\newcommand{\bmw}{\bm{w}}

\newcommand{\bmlambda}{\boldsymbol{\lambda}}
\newcommand{\bmxi}{\boldsymbol{\xi}}
\newcommand{\bmzeta}{\boldsymbol{\zeta}}

\newcommand{\be}{\begin{eqnarray}}
\newcommand{\ee}{\end{eqnarray}}

\newcommand{\w}{\ensuremath{\omega}}
\newcommand{\C}{\mathcal{C}}
\newcommand{\h}{\mathcal{H}}

\newcommand{\rr}{{\mathbf r}}
\newcommand{\pp}{{\mathbf p}}
\newcommand{\ag}{{\mathbf a}}
\newcommand{\vg}{{\mathbf v}}

\newcommand{\Rg}{{\mathbf R}}
\newcommand{\Yg}{{\mathbf Y}}
\newcommand{\zg}{{\mathbf 0}}

\newcommand*{\multint}{\ensuremath{\int \!\!\!\!\:\int}}
\newcommand*\diff{\mathop{}\!\mathrm{d}}
\providecommand*{\deriv}[3][]{\frac{\diff^{#1}#2}{\diff #3^{#1}}}
\providecommand*{\pderiv}[3][]{\frac{\partial^{#1}#2}{\partial #3^{#1}}}
\providecommand*{\eu}{\ensuremath{\mathrm{e}}}
\providecommand*{\iu}{\ensuremath{\mathrm{i}}}

\begin{document}


\begin{frontmatter}

\title{Heavy quark bound states in a quark-gluon plasma: dissociation and recombination}

\author{Jean-Paul Blaizot}
\address{Institut de Physique Th\'{e}orique (IPhT),  CNRS/URA2306, CEA Saclay,
F-91191 Gif-sur-Yvette, France}%
\author{Davide De Boni}
\address{Department of Physics, Swansea University, Swansea SA2 8PP, Wales, United Kingdom}
\author{Pietro Faccioli, Giovanni Garberoglio}%
\address{Dipartimento di Fisica  Universit\'a degli Studi di Trento and\\
Trento Institute for Fundamental Physics and Applications (INFN-TIFPA), \\Via Sommarive 14, Povo (Trento) 38123, Italy}%


\begin{abstract}
We present a comprehensive approach to the dynamics of heavy quarks in a quark gluon  plasma, including the possibility of bound state formation and dissociation. In this exploratory paper, we restrict ourselves to the case of an Abelian plasma, but the extension of the techniques used to the non Abelian case is straightforward. A chain of well defined approximations leads eventually to a generalized Langevin equation, where the force and the noise terms are determined from a correlation function of the equilibrium plasma, and depend explicitly on the configuration of the heavy quarks. We solve the Langevin equation for various initial conditions, various numbers of heavy quark-antiquark pairs, and various temperatures of the plasma. Results of simulations  illustrate various expected phenomena: dissociation of bound states as a result of combined effects of screening of the potential and collisions with the plasma constituent, formation of bound pairs (recombination) that occurs when enough heavy quarks are present in the system.  

\end{abstract}

\end{frontmatter}

\section{Introduction}

Heavy quarks produced in ultra-relativistic heavy ion collisions are interesting for a variety of reasons. They are created through hard processes taking place in small space time regions, at the very beginning of the collisions, and their abundances remain essentially frozen for the entire duration of the collisions.  Thus heavy quarks  can be used to diagnose the properties of the matter they cross before hadronizing. Heavy quarks can make bound states, such as the $J/\Psi$ or the $\Upsilon$ mesons, and the formation of such bound states can be strongly affected by the presence of a high temperature quark-gluon plasma. If the temperature  of such a plasma is high enough, the binding forces can be screened at very short distance, hindering  bound state formation, as was originally advocated in \cite{Matsui:1986dk}.  It was also argued that the presence of hot matter in the vicinity of a $J/\Psi$ meson could produce an observable mass shift \cite{Hashimoto:1986nn}. 
Note that other mechanisms,  besides  screening, involving in particular the collisions with the plasma constituents, can  lead to bound state dissociation. This is so for instance of the analog of photo-dissociation, namely gluo-dissociation \cite{Shuryak:1978ij,Peskin:1979va,Kharzeev:1994pz} (see also \cite{Brambilla:2013dpa} for a recent study).

In situations where heavy quarks are abundantly produced, an excess of  bound states could occur, due for instance to an enhanced recombination of $c\bar c$ pairs into $J/\Psi$ mesons at hadronization. 
Such a possibility was pointed  out early on  in Ref. \cite{MatsuiLBL} (see also\cite{Svetitsky:1988wv}). Amusingly, the concern at that time was that the recombination mechanism could spoil the proposed signature of quark-gluon plasma, by hiding the expected $J/\Psi$ suppression. Recombination was studied systematically using kinetic equations, viewing bound state formation and dissociation as a chemical reaction \cite{Thews:2000rj,Thews:2005vj}. A more extreme point of view is that  bound states never truly form in a plasma (we shall come back to this important point shortly), but only when matter hadronizes, recombination being then treated as a 
statistical process \cite{BraunMunzinger:2000px} (see also \cite{Andronic:2003zv} and \cite{Grandchamp:2001pf}).  Note that, in contrast to the initial worries, evidence for recombination would indicate that heavy quarks roam over long distances through the quark gluon plasma before recombining, thereby revealing a rather direct picture of a deconfined medium.

It turns out that  the predicted phenomenon of $J/\Psi$ suppression was observed experimentally in the first heavy ion collisions at the SPS, and later at RHIC (the reviews \cite{Rapp:2008tf,Kluberg:2009wc}  include a discussion of experimental results from SPS to RHIC).
More recently, evidence was obtained at the LHC for  sequential dissociation of the $\Upsilon$ bound states, with the less bound states being more suppressed than the most tightly bound ones
 \cite{Chatrchyan:2012lxa}. Some evidence for recombination was also presented by the  ALICE experiment \cite{Abelev:2013ila}. The interpretation of the data remains as of today uncertain for a variety of reasons, most of which having to do with the production mechanism of the heavy quark bound states (involving issues related to structure functions, shadowing, etc), the heavy ion reaction dynamics, etc, all aspects that are beyond the scope of the present paper.  However, the quality of the recent data, and the potential of upcoming experiments, provide strong motivation for further theoretical efforts.

Many investigations concern the fate of the $Q\bar Q$ bound state immersed in a quark gluon plasma in equilibrium at some temperature $T$. Such studies where initiated in \cite{Karsch:1987pv}  with the determination of the stationary states of a Schr\"odinger equation, with a temperature dependent potential that accounts for  screening and the disappearance of the string tension at high temperature. There has been discussion on the ambiguity in the choice of the appropriate potential (free energy versus internal energy) and how to relate it to quantities calculable on the lattice. A review of such potential models can be found in \cite{Mocsy:2013syh}. A somewhat similar line of research concerns the calculation of the spectral functions of charmonium states on the lattice. While such calculations have the virtue of being first principle calculations within Quantum Chromodynamics (QCD), they suffer from a large uncertainty in the reconstruction of the spectral function through the maximum entropy technique (for a recent review of lattice calculation at finite temperature, including a discussion of this issue, see e.g. \cite{Petreczky:2012rq}). 
Going somewhat beyond the Schr\"odinger picture, the in-medium T-matrix approach \cite{Mannarelli:2005pz} can encompass many effects beyond the screening of the potential;  it can in principle deal with dissociation reactions, changes in thresholds related to shifts of masses, coupled channels, etc. 

A recent progress in the direction of a more complete dynamical approach based on first principles was initiated in \cite{Laine:2006ns}. There one calculates a correlator that is directly related to an observable, the rate of  dilepton production, and derive the Schr\"odinger equation obeyed by this correlator. A remarkable feature of this equation is that the potential that enters it has an imaginary part that   reflects the effects of the collisions that the heavy quarks suffer with the plasma constituents. The origin of this complex potential was further studied in \cite{Beraudo:2007ky}, and also in the context of the non relativistic heavy quark effective theory in \cite{Brambilla:2008cx,Brambilla:2010vq}.

However, an important issue rarely addressed in the approaches that we have mentioned is that 
 the process of the bound state formation is not instantaneous: heavy quarks start to interact with the plasma while the correlations that could eventually lead to a bound state build up (see e.g. \cite{Svetitsky:1987gq} or \cite{Blaizot:1988ec} for early discussions of this issue).  This is an important feature that should be taken into account when trying to get a complete dynamical picture. In short, it is clear that the heavy quarks will suffer collisions with the constituents of the surrounding plasma as soon as they are created, and the real issue is whether they will still form a bound state when the plasma has cooled down, not whether the bound state will ``survive''. \\

The goal of the present paper is  to address this and other issues, by developing a comprehensive approach of the entire dynamics of the heavy quarks, including the possibilities for bound state formation and dissociation. We shall do that trying to stay as close as possible to first principles, and using a chain  of well defined (in some cases well controlled) approximations. The main objective is to get a global view of the dynamics of heavy quarks. As we proceed we shall recover some of the  many pieces that have been addressed separately in some of the works mentioned above. The approach builds up, extends, and to some extent completes, previous works by some of the authors \cite{Beraudo:2007ky,Beraudo:2010tw}. It is similar in spirit to analogous recent efforts using the language of open quantum system \cite{Akamatsu:2011se,Akamatsu:2012vt,Akamatsu:2014qsa} (see also \cite{Young:2010jq,Borghini:2011ms}). We shall be led eventually to formulate, at the end of our approximation chain, a generalized Langevin equation, and in that respect, our work bears similarities with previous studies using Langevin dynamics for heavy quarks  \cite{Moore:2004tg,Young:2008he,Young:2009tj}. The present  work goes beyond such studies by taking into account the dependence of the noise term  on the configuration of the heavy quarks at each time step. This is an important aspect of the dynamics, but it makes the Langevin equation more difficult to solve. In devising suitable techniques to handle it, we were inspired by similar problems in other fields of physics, 
  in particular by techniques used in soft matter physics \cite{Schneider13}.

  In this exploratory work we focus on the general issue of the formation of bound sates of heavy particles in a thermal bath of light particles.   The  paper is organized as follows. In the next section we summarize the general formalism that we are using. We treat the heavy particles as non relativistic, and the plasma particles as relativistic. We ignore the   specifics of QCD interactions, and for instance the change of their nature (from attractive to repulsive) depending on the color state (singlet or octet) the quark-antiquark pairs are in.
Only Coulomb interactions are retained. The goal of this section is to write the probability for a collection of quarks started at some positions at initial time to be found at a later time at some other positions. This is formulated in terms of a path integral from which we derive  an effective theory for the heavy quarks, usually referred to in such a context as the ``influence functional''. The main approximation in the elimination of the plasma degrees of freedom is a weak coupling ansatz that allows us to ignore  non linear interactions among the gluons. With this approximation, the effective action for the heavy quarks takes the simple form of an action quadratic in the charge density of the heavy quarks. The whole information about the plasma is contained in a 2-point function.  The content of the functional is analyzed in section 3 where we study the infinite mass limit of a related object, the correlator of a heavy quark-antiquark pair. In this case,  the plasma 2-point function reduces to a complex potential whose real part describes the screening phenomenon, while the imaginary part takes into account the effects of the collisions. The infinite mass limit is the leading order of a systematic approximation, the low frequency approximation, that we use to calculate the influence functional  in the following section. The  low frequency approximation  exploits the fact that the mass of the heavy quark is large, and can be viewed as an expansion in terms of the velocity of the heavy particles. The leading terms yield a generalized Langevin equation with a multiplicative (position dependent) noise. The ingredients of the Langevin equation are related to the real and imaginary part of the potential. In Sect.~5 we present results of a set of simulations that we have carried out for the case of 2, 10 or 50 pairs of particles. The various aspects of heavy quark propagation in a quark gluon plasma are illustrated, including the competing phenomena of dissociation and recombination. The last section contains a brief summary.  

\section{General setup}

The general problem that we are addressing is that of the dynamics of a collection of heavy charged particles in a thermalized bath of light particles with which they interact. Although our ultimate goal is to treat QCD interactions, in this exploratory paper we focus on the case of Abelian interactions, that is, strictly speaking our picture applies directly to electromagnetic interactions. Still we shall use the language of QCD, and call the heavy charged particles quarks (with positive unit charge) or antiquarks (with negative unit charge). Similarly the plasma in which the heavy particles move will consist of massless charged particles, referred to as light quarks, and the photons they exchange will be called gluons. We hope this abuse of language will not cause confusion. It is motivated by the fact that the simulations that we shall present in this paper involve parameters that are adjusted so as to lead to orders of magnitude that are comparable to what we expect for heavy quarks in a quark-gluon plasma (in particular we use the strong coupling constant $\alpha_s\approx 0.3$, not the electromagnetic one $\alpha\approx1/137$). Most of the effects that we want to study would occur already in Abelian plasmas. Specific effects due to non Abelian  color interactions will be discussed in forthcoming publications.

In this section we outline the general formalism that we use. The present approach builds on, and extends,  that developed by some of us in Refs.~\cite{Beraudo:2007ky,Beraudo:2010tw}, and the notation used here is close to that of these papers. A related effort was undertaken recently in Refs.~\cite{Akamatsu:2011se,Akamatsu:2012vt,Akamatsu:2014qsa}  which directly address the case of non Abelian plasmas. Our  goal is to obtain an effective theory for the heavy quarks, by eliminating the plasma degrees of freedom. This will be achieved by exploiting the fact that the heavy quarks behave as non relativistic particles, whose number is conserved as they interact with the quark-gluon plasma, and whose mass is large compared to the scales that characterize the plasma dynamics. Ultimately, the plasma properties will enter the effective theory only through specific correlation functions that are, in leading order, simply related to the potential whose real part describes the screening phonemenon, and the imaginary part the effects of collisions.\\

The heavy quarks, as we just mentioned,  are treated as massive non relativistic particles. When they thermalize, their typical wavelength, $\lambda\sim 1/\sqrt{MT}$ with $M$ is the mass and $T$ the temperature,  is small compared to the inter particle distance of the plasma particles, $1/T$. This suggests that the dynamics of the heavy particles is to a large extent classical, and indeed the approximation scheme that we shall present will lead us to a semi-classical description.  The heavy particles interact among themselves, and with the charged plasma constituents via Coulomb interactions. We neglect magnetic interactions, which are suppressed by powers of the velocity, or the inverse mass of the heavy particles. Such restriction do not apply a priori to the light particles, but since heavy quarks will not excite magnetic modes, we ignore these altogether. Thus the plasma is modeled  by massless quarks and antiquarks, interacting also with Coulomb interactions.  As we shall see, the dynamics of the plasma is then characterized by a unique energy scale which is the screening mass $m_D$. In order to get sensible orders of magnitude later in our simulations, we choose this to be given by its perturbative value for a two flavor quark-gluon plasma, i.e., $m_D^2=(4/3) g^2 T^2$ where $g$ is the gauge coupling. We shall assume throughout this paper  that $m_D< T\ll M\,$. 

In Coulomb gauge $\nab \cdot \A=0$,  the Hamiltonian of the system reads
\be\label{eq:hamiltonian}
H &=& \frac{1}{2M}\sum_{j=1}^N\left(\pp_j^2 + \bar\pp_j^2\right) + \int\diff{\x}~{\psi}^\dagger(\x)~\left(\frac{\balpha\cdot \nab}{i}+m\gamma_0\right)~\psi(\x) +\nn\\
&+& \frac{1}{2}\multint\diff{\x}\diff{\y}~\rho_{_{\scriptsize \mbox{tot}}}(\x)K(\x-\y)\rho_{_{\scriptsize \mbox{tot}}}(\y),
\ee
where $\alpha^i=\gamma_0 \gamma^i$ is a Dirac matrix, and  $\rho_{_{\scriptsize \mbox{tot}}}=\rho+\rho_\psi$ is the total charge density, with 
\be\label{rho}
\rho(\x) = g\,\sum_{j=1}^N\left[ \delta(\x- \q_j) -\delta(\x-\bar\q_j)\right],
\ee 
the charge density of the heavy quarks and antiquarks,  and
\be
\rho_\psi(\x)=g\,\psi^{\dagger}(\x)\psi(\x)
\ee
the density of the charged plasma particles (here, charged light quarks and anitquarks). The plasma is supposed to be electrically neutral, that is, it contains the same number of light quarks and antiquarks. It is useful to rewrite the Hamiltonian (\ref{eq:hamiltonian}) by separating its various contributions as follows
\beq
H=H_Q+H_1+H_{\rm pl},
\eeq
with $H_Q$ describing the dynamics of the heavy quarks in the absence of the plasma, 
\beq\label{HQham}
H_Q= \frac{1}{2M}\sum_{j=1}^N\left(\pp_j^2 + \bar\pp_j^2\right)+ \frac{1}{2}\multint\diff{\x}\diff{\y}\,\rho(\x)K(\x-\y)\rho(\y),
\eeq
$H_1$ the Hamiltonian coupling the heavy quarks to the plasma charged particles
\beq
H_1=\multint\diff{\x}\diff{\y}\,\rho(\x)K(\x-\y)\rho_\psi(\y),
\eeq
and $H_{\rm pl}$ the Hamiltonian of the plasma in the absence of the heavy quarks
\beq
H_{\rm pl}= \int\diff{\x}\,{\psi}^\dagger(\x)\left(\frac{\balpha\cdot \nab}{i}+m\gamma_0\right)\psi(\x) +
 \frac{1}{2}\multint\diff{\x}\diff{\y}\,\rho_\psi(\x)K(\x-\y)\rho_\psi(\y),
 \eeq

We represent the heavy particles in first quantization (they are non relativistic particles whose number is conserved), while the light particles are represented by the fermion fields $\psi(\x)$ and $\psi^\dagger(\x)$.  Note that the interaction term in Eq.~(\ref{HQham})  contains contributions of self interactions. Such terms will not contribute in the final equations that enter our simulations\footnote{They play a role in the real part of the potential to be discussed in Sect.~\ref{influence functional}.}. We call $\q_j$ and $\bar\q_j$, with $j=1,\cdots,N$, the coordinates of, respectively, the heavy quarks and antiquarks, and $\p_j$, $\bar\p_j$ the corresponding momenta. We denote collectively these coordinates   by a $2N$ dimensional vector $\Q=\left(\q_1,\cdots,\q_N,\bar\q_1,\cdots,\bar\q_N\right)\,$, and often refer to $\Q$ as a configuration. The last term in Eq.~(\ref{eq:hamiltonian}) is the total Coulomb energy, with 
\beq
K(\x-\y)= \frac{1}{4\pi|\x-\y|},\qquad -\nab^2_x K(x-y)=\delta(x-y).
\eeq

 We are interested in the probability $P(\Q_f, t_f|\Q_i, t_i) $ to find the heavy particles in a configuration $\Q_f$ at time $t_f$, given that they are in a configuration $\Q_i$ at time $t_i$. This  probability may be written in terms of the density matrix of the entire system (the plasma and the heavy particles). Let ${\cal D}$ be this density matrix. At time $t_i$ we assume that the heavy quarks have not yet interacted with the plasma, so that the density matrix takes the factorized form ${\cal D}(0)={\cal D}_Q^{(i)}\otimes {\cal D}_{\rm pl}^{(i)}$, with ${\cal D}_Q^{(i)}=|\Q_i\rangle\langle \Q_i|$, a projector on the configuration $\Q_i$,  and ${\cal D}_{\rm pl}^{(i)}=\frac{1}{Z_{\rm pl}}\rme^{-\beta H_{\rm pl}}$ is the density matrix of  the plasma in thermal equilibrium at temperature $T=1/\beta$, with $Z_{\rm pl}=\exp(\beta F_{\rm pl})$ the partition function of the plasma and $F_{\rm pl}$ its free energy. The density matrix at time $t$ is given by ${\cal D}(t)=\rme^{-iHt}{\cal D}(0)\rme^{iHt}$, and the looked for probability can be written in the form
\beq
P(\Q_f, t_f|\Q_i, t_i)&=&{\rm Tr} \left\{\left[ |\Q_f\rangle\langle \Q_f|\otimes {    \mathbb{I}}\right] {\cal D}(t)\right\}\nn\\
&=& \sum_{n,m} \frac{  \rme^{-\beta E^{\rm pl}_m}   }{Z_{\rm pl}} \langle \Q_i,m|\rme^{iHt}|\Q_f,n\rangle\langle \Q_f,n|\rme^{-iHt}|\Q_i,m\rangle\nn\\
&=&\sum_{n,m} \frac{  \rme^{-\beta E^{\rm pl}_m}   }{Z_{\rm pl}}\left|\langle \Q_f,n|\rme^{-iHt}|\Q_i,m\rangle\right|^2,
\eeq
where we have set $t\equiv t_f-t_i$. In order to trace out the degrees of freedom of the plasma, as implied by the formula above, it is convenient to rewrite this expression in terms of path integrals. 
\\

We shall do so in steps, in order to identify  the main components of the formalism. Let us first consider the simple case where the heavy particles interact only with an external potential $A_0(\x)$, with a Hamiltonian $H_1=g\int_{\x} \rho(\x)A_0(\x)$. In this case\footnote{As we shall see shortly  this case is relevant for the general discussion.}
\beq\label{probability0}
P(\Q_f, t_f|\Q_i, t_i) =\left| \langle\Q_f,t_f|\Q_i,t_i\rangle\right|^2. 
\eeq
and 
the  probability amplitude $\langle \Q_f,t_f|\Q_i,t_i \rangle$ is given by a Feynman path integral
\beq
\langle \Q_f,t_f|\Q_i,t_i \rangle&=&\int_{\Q_i}^{\Q_f} D\Q\;\eu^{i(S_0[\Q]+S_1[\Q,A_0])},
\eeq
where the paths $\Q(t)$ satisfy $\Q(t_i)=\Q_i$ and $\Q(t_f)=\Q_f$. The actions $S_0$ and $S_1$ are given by
\beq\label{S0}
S_0[\Q]&=&\frac{M}{2}\sum_{j=1}^N\int_{t_i}^{t_f} \rmd t\left(\dot \q^2_j +\dot{\bar\q}^2_j \right),\\ \label{S1} S_1[\Q, A_0] &=&-\int_{t_i}^{t_f}\rmd t\,\int\rmd^3\x\,\rho(x) A_0(x),
\eeq
where we have  set $x=(t,\x)$, and $\rho(x)$ is the charge density of the heavy particles
\beq
\rho(x)=\sum_{j=1}^N g\left[  \delta(\x-\q_j(t)-\delta(\x-\bar\q_j(t)  \right],
\eeq
so that $S_1$ can also be written as
\beq
S_1[\Q,A_0]=-g\sum_{j=1}^N\int_{t_i}^{t_f}\!\!\!\rmd t\,\left[A_0(\q_j(t)) -A_0(\bar\q_j(t))\right].
\eeq

\begin{figure}[t!]
\begin{center}
\includegraphics[width=12cm]{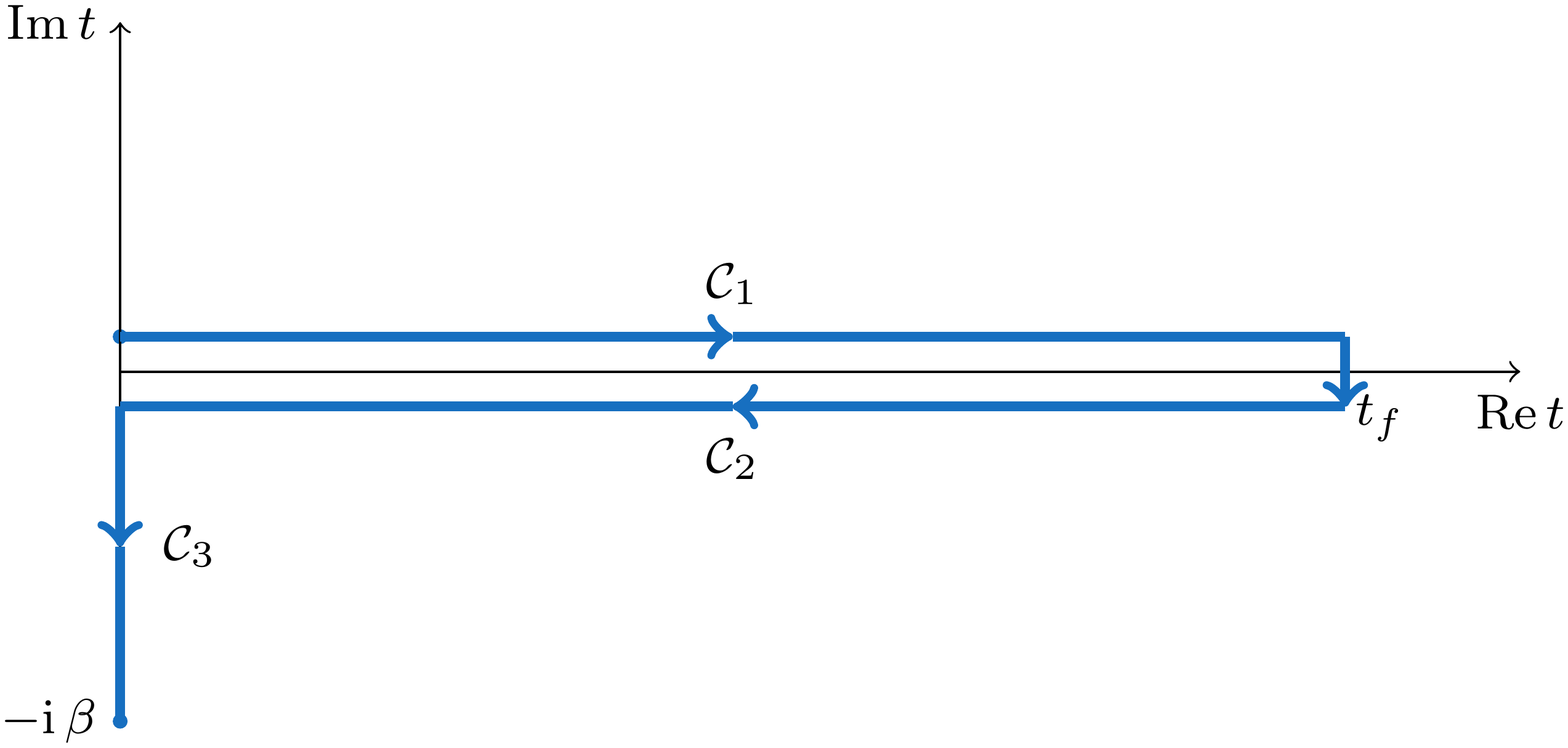}
\caption{The Keldysh contour $\C$, with its different branches.}
\label{keldysh}
\end{center}
\end{figure}

The probability (\ref{probability0}) can be represented by a very similar formula, by using the closed-time path formalism. We introduce a contour in the complex time plane, as illustrated in Fig.~\ref{keldysh}, and consider the coordinates $\{\q_i,\bar\q_i\}$ as functions of the complex time $t_{\scriptsize{\cal C}}$ running along the contour. Alternatively, we may keep time real, but duplicate the coordinates,  denoting  by $\Q_1=\{ \q_{i,1},\bar\q_{i,1}  \}$ and $\Q_2=\{ \q_{i,2},\bar\q_{i,2}  \}$ the coordinates of the heavy particles living respectively on the upper branch (${\cal C}_1$, corresponding to the amplitude) and the lower branch (${\cal C}_2$, corresponding to the complex conjugate amplitude) of the contour. We can then write
\beq\label{probability1}
P(\Q_f, t_f|\Q_i, t_i) =\int_{\cal C} D\Q\;\eu^{i(S_0[\Q]+S_1[\Q,A_0])},
\eeq
where the actions $S_0$ and $S_1$ are given by the formulae (\ref{S0},\ref{S1}) in which the time integrations are replaced by integrations along the Schwinger-Keldysh contour. Thus, for instance, $S_0$ is given by
\beq
S_0[\Q]=\frac{M}{2}\sum_{j=1}^N\int_{\cal C} \rmd t^{_\C}\left(\dot \q^2_{j} +\dot{\bar\q}^2_{j} \right),
\eeq
where the time coordinate $t^{_\C}$ runs along the contour ${\cal C}$, that is, from $t_i$ to $t_f$ slightly above the real time axis, and returns from $t_f$ to $t_i$ slightly below it. Thus,
\beq
 \int_{\cal C} \rmd t^{_\C}\,\dot \q^2_{j}=\int_{t_i+i\eta}^{t_f+i\eta} \rmd t^{_\C}\,\dot\q_{j}^2+\int_{t_f-i\eta}^{t_i-i\eta} \rmd t^{_\C}\,\dot\q_{j}^2=\int_{t_i}^{t_f} \rmd t \,(\dot\q_{j,1}^2-\dot\q_{j,2}^2), 
\eeq
where in the last step we have duplicated the coordinate $\q_j(t)$ as discussed above. The last  expression appears naturally in the action when one multiplies the probability amplitude by its complex conjugate in order to build the probability  (\ref{probability1}), with $\q_{j,1}(t)$ labelling the path in the amplitude and $\q_{j,2}(t)$ the path in the complex conjugate amplitude.\\
\\

It is straightforward to extend the formula (\ref{probability1}) to include the interactions among the heavy particles and with the light plasma constituents.  We have (see e.g. \cite{calzetta} or \cite{Kleinert})
\be\label{PI2}
P(\Q_f, t_f|\Q_i, t_i) = \int_{\cal C} D\Q \int_{\cal C} D({\bar\psi,\psi})~\eu^{\iu\,S[\Q, \psi, \bar{\psi}]}\,, 
\ee
where the contour now includes a vertical piece, ${\cal C}_3$ corresponding to the thermal average of the plasma degrees of freedom at the initial time (i.e., the trace over the equilibrium density matrix of the plasma). Accordingly, the  fermionic fields in Eq.~(\ref{PI2}) obey anti-periodic boundary conditions on ${\cal C}_3$, $\psi(0,\x)=-\psi(-\iu \beta,\x)$, $\overline{\psi}(0,\x)=-\overline{\psi}(-\iu \beta,\x)$. The action $S[\Q, \psi, \bar{\psi}]$ is given by
\be\nn
&&S[\Q, \psi, \bar{\psi}] =S_0[\Q]+ \int_{\cal C} \rmd^4 x~\bar{\psi}(x)(~\iu\gamma^\mu \partial_\mu-m ~)\psi(x)\\
&&\qquad\qquad\qquad\qquad- \frac{1}{2}~\multint_{\cal C} \rmd^4 x\,\rmd^4 y~\rho_{_{\scriptsize \mbox{tot}}}(x)K(x-y)\rho_{_{\scriptsize \mbox{tot}}}(y)\:,
\ee
where  $K(x-y)=\delta(t_x-t_y)K(\x-\y)$ represents  the (instantaneous) Coulomb interaction, 
and $\rho_{\rm tot}$ is the total charge density.
It is important to stress that the heavy particles do not take part in the  thermal average, and consequently they do not propagate along the imaginary time sector of the Keldysh contour\footnote{Note that we use the notation $\int_{\cal C}$ to denote either a path integrals where the paths are defined on the contour, as in $\int_{\cal C}D\Q$, or an ordinary integral, as in $ \int_{\cal C} \rmd t^{_\C}$ where the time variable $t^{_\C}$ lives on the contour.}. We may take $\rho(t=-i\tau,\x)=0$, with $0<\tau\leq\beta$.\\

The next step consists in eliminating the light fermion field in favor of a Coulomb potential $A_0$. To this end, we use the formal identity\footnote{We follow closely here the approximation scheme developed  in Ref.~\cite{Beraudo:2010tw}}: 
\be\label{formalidentity1}
\exp\left[-\frac{\iu}{2}\,\rho_{\rm tot}\cdot K\cdot \rho_{\rm tot}\right]=\mathcal{N}\int_{\cal C} D A_0~ \exp\left[\frac{\iu}{2}\,A_0\cdot K^{-1} \cdot A_0-\iu\,A_0\cdot\rho_{\rm tot}\right],\nn\\
\ee
where $\mathcal{N} \sim \left(\mbox{det}\left[\nabla^2\right]\right)^\frac{1}{2} $  is a normalization constant, $K^{-1}(\x-\y)=-\delta(\x-\y)\nab_{\y}^2$, and we use a matrix notation to simplify the formulae, e.g., 
\beq
\rho\cdot K\cdot \rho=\multint_{\cal C} \rmd^4 x\,\rmd^4 y\, \rho(x) K(x,y)\rho(y).
\eeq
When using this identity, it is important to remember that part of the $A_0$ potential is the Coulomb field created by the heavy particles. We shall call  $A_0^{\rm cl}$ this contribution, and write accordingly $A_0=A_0^{\rm cl}+\tilde A_0$. By definition, we have
\beq
-\nab^2 A_0^{\rm cl}(\x)=\rho(\x), \qquad A_0^{\rm cl}(\x)=\int\rmd\y\, K(\x-\y)\rho(\y).
\eeq
The integration over $A_0$ is then truly an integration over the field $\tilde A_0$, and this field satisfies the imaginary time (KMS)  periodic boundary condition $\tilde A_0(0,\x)=\tilde A_0(-\iu\,\beta,\x)$. 
We could take this explicitly into account by performing a shift of integration variables. It is more convenient not to do so, provided that we remember that $A_0^{\rm cl}$ plays the role of a constant in the functional integral, in particular the result of such integration will depend on $A_0^{\rm cl}$.

Using the identity above, and remembering that $\rho_{\rm tot} =\rho+g\,\psi^\dagger \psi$, we can perform the Gaussian integrals over the light fermion fields
\be
&&\int D (\bar{\psi},\psi)~\exp\left[\iu\int_{_{\scriptsize \C}}\!\!\!\diff{x}~\bar{\psi}(x)(\iu\gamma^\mu \partial_\mu-m-g\gamma^0A_0(x))\psi(x)\right] \nn\\
&&\qquad\qquad\qquad\qquad\qquad= \exp\left[\mbox{Tr}\,\,\ln \left[\iu\gamma^\mu\partial_\mu - m -  g \gamma^0 A_0\right]\right].
\ee
The probability (\ref{PI2}) takes then the form  
\beq
P(\Q_f, t_f|\Q_i, t_i) = \int_{\cal C} D\Q \int_{\cal C} D A_0~\rme^{\iu\,\left(  S_0[\Q]+S_1[\Q, A_0]+S_2[\Q, A_0]\right)}, 
\eeq
where $S_1$ is given  by Eq.~(\ref{S1}) and 
\beq
S_2[\Q,A_0] = -\frac{1}{2}\int_{\cal C}\diff{x}\left(\,A_0(x)\nabla^2 A_0(x)\,\right) - \iu~\mbox{Tr}\,\ln \left[~\iu\gamma^\mu  \partial_\mu - m -  g \gamma^0 A_0(x)~\right].\nn
\eeq
The dependence of $S_2$ on $\Q$ arises from the dependence of $A_0^{\rm cl}$ on the positions of the heavy particles.

It is convenient to rewrite the integral over $A_0$ as the exponential of an effective action, the so-called Feynman-Vernon (FV) influence functional \cite{FV63}:
\be\label{FV}
\eu^{\iu \Phi[\Q,A_0^{\rm cl}]} =  \int \D A_0\,\eu^ {-\iu\int_{\cal C}\rmd^4x\,\rho(x)A_0(x)}\, \eu^{\iu S_2[A_0]}.
\ee
The exponential of the FV functional is the thermal average over the $A_0$ fluctuations of the exponential factor that contains the linear interaction $\rho A_0$ between the heavy particles, with charge density $\rho$, and the total Coulomb field $A_0$. This particular structure is a consequence of the fact that the heavy quark is linearly coupled to the total field $A_0$. So far, no approximation has been made (within the present Abelian context). We shall now introduce the main approximation of the whole approach, that consists in neglecting the non linear self-interactions of the $A_0$ field.\\

The action $S_2$  contains a non-local term describing the coupling between light quarks  and gluons, as well as the classical Coulomb interaction between the heavy particles.
Its expansion   in powers of $A_0$ gives rise to induced effective couplings to all orders in the coupling constant $g$. In order to be able to compute the influence  functional we need to introduce some approximations. We do so by retaining only the terms up to quadratic order in the coupling $g$, or equivalently in the field $A_0$. This is certainly an excellent approximation at truly weak coupling, like in electromagnetic plasmas. In the case of QCD, the validity of this weak coupling approximation may require further investigation. The main virtue of this approximation is to make  the path integral over $A_0$ calculable, since it becomes Gaussian. The influence functional  $\Phi[\Q]$ becomes 
\be\label{eq:phi}
\Phi[\Q] = \frac{1}{2}\multint_{\cal C}\,\rmd^4x\rmd^4y~\rho(x)\Delta_{_{\scriptsize \C}}(x-y)\rho(y)\:,
\ee
where 
\beq
\Delta_{\cal C}(x-y) = \iu\langle\, T_{\C} \left[ A_{0}(x) A_{0}(y)\,\right] \rangle
\eeq
 is the  longitudinal  gluon propagator (see next section for an explicit expression) defined on the contour, and obeying the  KMS conditions. Its inverse involves the   $1$-loop longitudinal photon self-energy $\Pi_{00}^{\C}$, also defined on the contour,
\be\label{Delta_C}
-\Delta_{_{\scriptsize \C}}^{-1}(x-y)=\delta_{_{\scriptsize \C}}(t_x^{_\C}-t_y^{_\C})~K^{-1}(\x-\y) + \Pi_{00}^{\C}(x-y), 
\ee
where $\delta_{_{\scriptsize \C}}(x-y)=\delta_{_{\scriptsize \C}}(t_x^{_\C}-t_y^{_\C})\,\delta(\x-\y)$.

It is convenient to write the influence functional using the duplicated fields, that is, we make the substitution 
\be\label{eq:notation}
\Q(t^{_\C}) &\to& (\Q_1(t), \Q_2(t))\\
A_0(t^{_\C}, \x) &\to& (A_{0,1}(t, \x),A_{0,2}(t, \x))\nn,
\ee
where $t^{_\C}$ denotes the curvilinear abscissa parametrizing the Keldysh contour, while $t\in [t_i,t_f]$ denotes the physical time. The integration over the physical time is always from $t_i$ to $t_f\,$.  With this  notation $\Delta$ becomes a matrix, 
\be
\Delta_{ab}(t_x-t_y)=\Delta(t_x^{_\C}-t_y^{_\C})  \quad \mbox{with} \quad t_x^{_\C}\in\C_a\,,\,t_y^{_\C}\in\C_b\:,\quad a,b=1,2.
\ee
We have, explicitly, 
\be
&&\Delta_{11}(x,y)=\iu\langle\, T_{_\C} \left[ A_{0,1}(x) A_{0,1}(y)\,\right] \rangle=\iu\langle\, T \left[ A_{0}(x) A_{0}(y)\,\right] \rangle=\Delta(x,y),\nn\\
&&\Delta_{21}(x,y)=\iu\langle\, T_{_\C} \left[ A_{0,2}(x) A_{0,1}(y)\,\right] \rangle=\iu\langle\,  A_{0}(x) A_{0}(y)\, \rangle=\iu\Delta^> (x,y),\nn\\
&&\Delta_{12}(x,y)=\iu\langle\, T_{_\C} \left[ A_{0,1}(x) A_{0,2}(y)\,\right] \rangle=\iu\langle\,  A_{0 }(y) A_{0}(x)\, \rangle=\iu\Delta^< (x,y),\nn\\
&&\Delta_{22}(x-y)=\iu\langle\, T_{_\C} \left[ A_{0,2}(x) A_{0,2}(y)\,\right] \rangle=\iu\langle\, \tilde T \left[ A_{0}(x) A_{0}(y)\,\right] \rangle=\tilde \Delta (x,y).\label{Deltaab}
\ee
where $T$ and $\widetilde{T}$ denote respectively the time ordering and anti ordering, and $\langle\cdots\rangle$ is the thermal average. Using this notation, the phase in Eq.~(\ref{eq:phi}) becomes
\be\label{PHI2}
\Phi[\Q] = \frac{1}{2} \int_{t_i}^{t_f}\diff{t_x}\int_{t_i}^{t_f}\diff{t_y}\int\diff{\x}\diff{\y}~(-1)^{a+b}\rho_a(t_x,\x)~\Delta_{ab}(t_x-t_y,\x-\y)~\rho_b(t_y,\y).\nn\\
\ee
 Note that we integrate all times the same way, i.e., on the real time axis from $t_i$ to $t_f$, so that the off diagonal terms pick up a minus sign. Note also that in the right hand sides of Eqs.~(\ref{Deltaab}), we have introduced the notation $\Delta$ (without subscripts) to denote the time-ordered real time propagator. Other useful relations are $\Delta=\Delta^R+\iu\Delta^<$, $\tilde \Delta=-\Delta^A+\iu\Delta^<$, where $\Delta^R$ and $\Delta^A$ denote respectively the retarded and the advanced propagators.  \\

At this point the probability (\ref{probability1}) is written as the following path integral
\beq\label{probability1b}
P(\Q_f, t_f|\Q_i, t_i) =\int_{\cal C} D\Q\;\eu^{iS_0[\Q]}\, \eu^{i\Phi[\Q]},
\eeq
with $\Phi[\Q]$  given by Eq.~(\ref{PHI2}) above. The plasma degrees of freedom have been eliminated, the plasma properties entering the calculation of $\Phi[\Q]$ solely through the contour propagator $\Delta_{ab}(t_x-t_y,\x-\y)$ that plays the role of an effective interaction between the heavy quarks. The problem of calculating the probability $P(\Q_f, t_f|\Q_i, t_i)$ has been reduced  to that of calculating an ``ordinary'' Feynman path integral. This remains however a difficult task, in particular when many heavy $Q\bar Q$ pairs are present  in the system, and we shall shortly proceed with further approximations. Before we do that, we shall consider in the next section a situation where the influence functional can be calculated exactly: this is the case where a single, infinitely massive, $Q \bar Q$ pair is present in the system.

\section { The influence functional and the complex potential}\label{influence functional}

We consider in this subsection the case of a single heavy $Q\bar Q$ pair. This will allow us to relate the influence functional to the complex potential that was first identified in this context in Ref.~\cite{Laine:2006ns}.
We denote here the coordinates of these heavy particles by $\Q=\{\r,\bar\r\}$, and we consider the correlator:
\beq
G^>(t_f,\Q_f|t_i,\Q_i)\equiv
\langle \psi_{\bar Q}(t_f,\bar\r_f)\psi_Q(t_f,\r_f)
\psi^\dagger_Q(t_i,\r_i)\psi^\dagger_{\bar Q}(t_i,\bar\r_i)\rangle,\label{eq:2parta}
\eeq
where the angular brackets represent the thermal average over the plasma particles (being understood that the states of the plasma do not contain any heavy quarks). This object enters directly the calculation of the heavy quarkonium spectral function, and for instance the calculation of dilepton emission rate \cite{Weldon:1990iw}.  As was shown in \cite{Beraudo:2007ky}, under the same approximations as those done presently, this correlator is given by 
\beq\label{eq:2parta2}
G^>(t_f,\Q_f|t_i,\Q_i)=\int_{\Q_i}^{\Q_f} D\Q \,\rme^{iS_0[\Q]}\rme^{i\Phi[\Q]}, 
\eeq
where $\Q$ lives on the upper part of the contour. The influence functional $\Phi$ has the same form as in Eq.~(\ref{PHI2}), that is,  
\be\label{PHI2b}
\Phi[\Q] = \frac{1}{2} \int_{t_i}^{t_f}\diff{t_x}\int_{t_i}^{t_f}\diff{t_y}\int\diff{\x}\diff{\y}~\rho(t_x,\x)~\Delta(t_x-t_y,\x-\y)~\rho(t_y,\y)\:,
\ee
but now all times are on the upper part of the contour, and here $\Delta=\Delta_{11}$ is the real time, time-ordered, propagator (see Eqs.~(\ref{Deltaab})). Recall that the density is $\rho(t_x,\x)=g\left[ \delta(\x-\r(t_x))-\delta(\x-\bar\r(t_x))\right]$, so that  the influence functional can be written as $\Phi_{_{QQ}}+\Phi_{_{\bar Q\bar Q}}+\Phi_{_{Q\bar Q}}$, with
\be
\Phi_{_{QQ}}[\Q] &=& \frac{g^2}{2} \int_{t_i}^{t_f}\diff{t_x}\int_{t_i}^{t_f}\diff{t_y}~\Delta(t_x-t_y,\r(t_x)-\r(t_y)),\label{PHI2c1}\\
\Phi_{_{\bar Q\bar Q}}[\Q] &=& \frac{g^2}{2} \int_{t_i}^{t_f}\diff{t_x}\int_{t_i}^{t_f}\diff{t_y}~\Delta(t_x-t_y,\bar \r(t_x)-\bar \r(t_y)),\label{PHI2c2}\\
\Phi_{_{Q\bar Q}}[\Q] &=& -{g^2} \int_{t_i}^{t_f}\diff{t_x}\int_{t_i}^{t_f}\diff{t_y}~ \Delta(t_x-t_y,\r(t_x)-\bar \r(t_y)).\label{PHI2c3}
\ee
In the last line we have used the fact that $\Delta(t, \x)$ is in fact a function of $|t|$ and $|\x|$ in order to rewrite $\Delta(t_x-t_y,\bar \r(t_x)-\r(t_y))$ as $\Delta(t_y-t_x, \r(t_y)-\bar\r(t_x))$, which coincides with the term already written after the interchange of the integration variables $t_x$ and $ t_y$. 
\\

The calculation is particularly simple in the infinite mass limit, where the paths are trivial (since infinitely heavy quarks do not move). In this case, the influence functional takes the form
\beq
\Phi[\Q] =-{g^2} \int_{t_i}^{t_f}\rmd t_x \int_{t_i}^{t_f}\rmd t_y \left[ \Delta(t_x-t_y,\r-\bar\r) -\Delta(t_x-t_y,0)   \right].
\eeq
At this point, it is convenient to  express $\Delta(t_x-t_y,\r-\bar\r)$ in terms of its Fourier transform
\beq
\Delta(t_x-t_y,\r-\bar\r)=\int\frac{\rmd \omega}{2\pi}\rme^{-\iu\omega (t_x-t_y)} \Delta(\omega,\r-\bar\r).
\eeq
This allows us to perform the time integrations 
\beq
 \int_{t_i}^{t_f}\rmd t_x \int_{t_i}^{t_f}\rmd t_y\, \rme^{-\iu\omega (t_x-t_y)} =\frac{2}{\omega^2}\left( 1-\cos(\omega(t_f-t_i)  \right),
\eeq
and obtain, after a further Fourier transform of the coordinates, 
\beq
\Phi[\Q] =2{g^2} \int\frac{\rmd \omega}{2\pi} \int\frac{\rmd \k}{(2\pi)^3} \frac{1-\cos(\omega t)}{\omega^2} \left[ \Delta(\w,\k)  -\rme^{\iu\k\cdot (\r-\bar\r)}\,\Delta(\w,\k)  \right],
\eeq
with $t=t_f-t_i$.

We are interested in the evolution of the heavy quarks on time scales that are large compared to the time scale that characterizes the dynamics of the plasma, and which is controlled by the inverse of the Debye mass, $m_D$. It is then useful to consider the large time limit of the expression above. This is easily obtained with the help of the relation   $(1-\cos(\omega t))/\omega^2\sim \pi t\delta(\omega)$ valid as $t\to\infty$ (i.e., $t\gg 1/\omega$). We get
\beq
\Phi[\Q] \simeq {g^2} (t_f-t_i)  \int\frac{\rmd \k}{(2\pi)^3} \left(  1-\rme^{\iu\k\cdot (\r-\bar\r)}\right)\Delta(0,\k).\nn\\
\eeq
Thus, at large time, the influence functional is dominated by the zero frequency part of the response function of the plasma.
As an alternative to the calculation done above, we could change integration variables, $t_x,t_y \to (t_x+t_y)/2, t_x-t_y$, and observe that when $t_f-t_i$ is large (compared to $m_D^{-1}$), on can integrate freely over $t_x-t_y$, which filters out the zero frequency component of the response.

By considering the equation of motion for the correlator (\ref{eq:2parta}) at large time, and its corresponding expression in terms of the influence functional, we interpret, following previous works,  the coefficient of $t=t_f-t_i$ in the influence functional  as a complex potential. That is, we set $\rme^{\iu\Phi}= \rme^{-\iu\,t {\cal V}}$. More precisely, remembering that $\Delta(0,\r)=\Delta^R(0,\r)+\iu\Delta^<(0,\r)$, we set 
\beq\label{complexpotential}
&&V(\r)= -\Delta^R(0,\r)=-\int\frac{d\k}{(2\pi)^3}\,e^{\iu \k\cdot
\r}\,\Delta^R(\omega=0,\k),\label{eq:repot}\\
&& W(\r)=-\Delta^<(0,\r)=-\int\frac{d\k}{(2\pi)^3}\,e^{\iu \k\cdot
\r}\,\Delta^<(0,\k),\label{eq:impot}
\eeq
so that 
\beq
{\cal V}(\r)=-g^2 [V(\r)-V_{\rm ren}(0)]-\iu\,g^2[ W(\r)-W(0)].
\eeq
where  the minus sign in front of $g^2$ is the same as in Eq.~(\ref{PHI2c3}) and reflects the fact that the $\r$ dependence of the potential describes interation between heavy quarks with opposite charges. At this point, we identify $\Delta$ with the real-time gluon propagator in Fourier space, at zero frequency. In the  hard thermal loop approximation \cite{HTL}, a suitable approximation in the present context, this is given  by (see e.g. \cite{Beraudo:2007ky})
\beq\label{eq:w0QED}
D_{L}(\omega=0,\k)=\frac{-1}{\k^2+m_D^2}+\iu\frac{\pi m_D^2 T}
{|\k|(\k^2+m_D^2)^2}, 
\eeq
which allows us to get an explicit expression for ${\cal V}(r)$,  a function of  $r\equiv|\r-\bar\r|$. The calculation of the integrals in Eqs.~(\ref{eq:repot}) and (\ref{eq:impot}) yields 
\beq
{\cal V}(r)=-\frac{g^2}{4\pi}m_D  -\frac{g^2}{4\pi} \frac{e^{-m_Dr}}{r}
-\iu\frac{g^2T}{4\pi}\phi(m_Dr)\,,
\eeq
where the first term is a self energy contribution, $V_{\rm ren}(0)$, that has been estimated by subtracting the corresponding vacuum part, thereby leaving the following thermal contribution  
\beq
\int_q\left(\frac{1}{\q^2+m_D^2}-\frac{1}{\q^2}\right)=-\frac{m_D}{4\pi}.
\eeq
Note that, as expected, the real part of the potential between
the quark and the anti-quark is attractive and screened. The
 imaginary part of the potential  originates from the collisions between the light fermions of the hot medium and  the heavy quarks. In fact, the quantity
 \beq
\Gamma=\frac{g^2T}{2} \int \frac{\rmd^3 \k}{(2\pi)^3} \frac{\pi m_D^2 T}{|\k|(\k^2+m_D^2)^2}
\eeq
is the rate of collisions between one heavy quark and the light quarks of the plasma. It may be identified with the damping factor associated with the propagation of one heavy quark in the plasma. The function
\beq
\phi(x)\equiv 2\int_0^\infty dz\frac{z}{(z^2+1)^2}
\left[1-\frac{\sin(zx)}{zx}\right],
\eeq 
which vanishes
for $x\!=\!0$ and increases monotonously, approaching unity as $x\!\to\!+\infty$. 
Thus,  the collisional damping rate is  most important when the heavy quarks are far apart: when this is so, the damping factor associated with the propagation of the heavy quark pair is just twice the damping factor of a single heavy quark.  When the heavy quark gets closer to the heavy quark, interference occurs that gradually suppresses the effect of collisions.  When the $Q\bar Q$ separation vanishes, this interference is completely destructive and kills the imaginary part: this is because when the $Q\bar Q$ separation is too small, the charge of the individual heavy quarks are not resolved by the typical fluctuation of the electric potential of the plasma.  The $Q\bar Q$ pair behaves then as a small, neutral, electric dipole, which propagates in the plasma without interacting. Note that   the behavior of the function $\phi(x)$ at small $x$ is singular:  $\phi(x)$ is continuous as $x\to 0$, but  it does not have a simple Taylor expansion, as can be seen from the logarithmic divergence of the coefficient of the term of order $x^2$. We come back to this issue in Sect.~(\ref{Sec:cutoff}).

Before we leave this section we should emphasize an important difference between the calculation that we have just presented of the correlator (\ref{eq:2parta2}), and that of the probability (\ref{probability1b}). Superficially, these quantities differ solely by the contour involved in the integration of the influence functional. In fact this change of contour is not innocent, and the two quantities are deeply different.   The correlator (\ref{eq:2parta}) may be interpreted as a probability amplitude to find the $Q\bar Q$ pairs in configuration $\Q_f$ at time $t_f$ together with the plasma in the same state as it is at time $t_i$, irrespective of what that state is. Because, during their  propagations, the heavy quarks mix with complicated configurations that involve plasma constituents, the amplitude    decays with increasing time, and this even when the quarks are infinitesimally heavy. This is the origin of the imaginary part of the potential, an imaginary part that would affect also the analog of the correlator (\ref{eq:2parta}) for a single quark \cite{Beraudo:2010tw}. The probability (\ref{probability1b}) addresses another question, namely the probability to find the heavy quarks in the configuration $\Q_f$, irrespective of the state of the plasma. In the limit of infinitely heavy quarks, one expects this probability to be proportional to $\delta(\Q_f-\Q_i)$ and this is indeed what we shall verify in the next section.

\section{Low frequency approximation and generalized Langevin equation}\label{low}

We now return to the calculation of the probability (\ref{probability1b}), and introduce an approximation, the low frequency approximation, that allows us to go beyond the infinite mass limit that we have considered in the previous subsection. Still, as we shall see, in this approximation, only the basic quantities that appear in the infinite mass limit will be needed, namely the real an the imaginary part of the potential. 

\subsection{The low frequency approximation}

The approximation  relies on the fact that the dynamics of the heavy fermions is much slower than the dynamics of the light fermions of the medium.  As we have recalled,   the typical frequency in, for instance, the time-ordered propagator\footnote{Since the spatial coordinates, or the three momenta, play no role in this discussion, we omit them to simplify the notation and denote the propagator $\Delta(\omega,\rr)$ or $\Delta(\omega,\k)$ simply by $\Delta(\omega)$.} $\Delta(\omega)$, is $m_D$.  Now, during a time $t$,  the heavy quark moves a typical distance $\sim \sqrt{t/M}$. For $t\sim m_D^{-1}$ this is a small distance compared to the size of the screening cloud, $\sim m_D^{-1}$: $m_D \sqrt{m_D^{-1}/M}\sim \sqrt{m_D/M}\ll 1\,$. Thus, over a time scale characteristic of the plasma collective dynamics,  the heavy quark positions are almost frozen (they are completely frozen in the limit $M\to\infty$). Said differently, the plasma dynamics looks very fast to the heavy quarks, and their interactions with plasma constituents are essentially instantaneous. In order to  observe a substantial motion of the heavy quarks, we need to look at the system over time scales that are large compared to $m_D^{-1}$.  In the calculation of the influence functional, we need therefore to allow for $t_x-t_y\gg m_D^{-1}$, or equivalently, in  Fourier space,   typical frequencies $\omega\ll m_D$. To get a systematic expansion, one  expands $\Delta(\omega)$ in powers of $\omega$ around $\omega=0$. In leading order this yields
\beq\label{lowfreqapprox}
\Delta(t_x-t_y)&=&\int\frac{\rmd \omega}{2\pi}\rme^{-\iu\omega (t_x-t_y)}\left[ \Delta(\omega=0)+\omega  \Delta^\prime(\omega=0)\right]\nn\\
&\simeq&\delta (t_x-t_y)\Delta(\omega=0)+\iu\frac{\rm d}{\rmd t_x}\delta (t_x-t_y)\Delta^\prime(\omega=0).
\eeq
This expansion shows indeed  that the heavy quarks interact with the medium via effectively instantaneous interactions. One recognizes in the first term of the expansion the infinite mass limit. The corrections implied by the second term will involve the velocities of the heavy quarks, as we shall see shortly.

We have identified in the previous subsection the zero frequency part of the time ordered propagator to the complex potential. Because of the relation obeyed by the various components of the propagator (see for instance \cite{Blaizot:2001nr}), the  derivative term $\Delta^\prime(\omega=0)$ is simply related to the imaginary part of the potential, as we now show. Indeed, the time ordered propagator can be written as  $\Delta(\omega)=\Delta^R(\omega)+\iu\Delta^<(\omega)$, where $\Delta^R(\omega)$ is the retarded propagator and $\Delta^<(\omega)$ has been defined above. The latter is related to the other function $\Delta^>(\omega)$ by the KMS relation, $\Delta^>(\omega)=\rme^{\beta\omega}\Delta^<(\omega)$, and the two functions allow us to reconstruct the spectral density $\rho(\omega)=\Delta^>(\omega)-\Delta^<(\omega)$.  From the last two equations, one easily establishes that $\Delta^<(\omega)=N(\omega)\rho(\omega)$, with $N(\omega)=1/(\rme^{\beta\omega}-1)$. From this relation, and using the fact that the spectral function is an odd function of $\omega$, it is easy to show that $\Delta^>(-\omega)=\Delta^<(\omega)$, so that, in particular, $\Delta^<(0)=\Delta^>(0)$. It follows then easily that
\beq
\left.\frac{\rmd \Delta^>}{\rmd \omega}\right|_{\omega=0}=-\left.\frac{\rmd \Delta^<}{\rmd \omega}\right|_{\omega=0},\qquad \left.\frac{\rmd \Delta^<}{\rmd \omega}\right|_{\omega=0}=-\frac{\beta}{2} \Delta^<(0).
\eeq
Furthermore, it is easily shown using the spectral representation of the retarded function, and again the fact that the spectral density is an odd function of $\omega$, that $\left.\rmd \Delta^R(\omega)/\rmd \omega\right|_{\omega=0}=0$. Therefore, 
\beq\label{derivDelta}
\left.\frac{\rm d\Delta(\omega)}{\rmd\omega}\right|_{\omega=0} = \iu\left.\frac{\rm d\Delta^<(\omega)}{\rmd\omega}\right|_{\omega=0}=- \iu\frac{\beta}{2} \Delta^<(\omega=0).
\eeq
Note finally that $\Delta_{22}(\omega)=-\Delta^A(\omega)+ \iu\Delta^<(\omega)$, where $\Delta^A(\omega)$ denotes the advanced propagator. At zero frequency, $\Delta^A(\omega=0)=\Delta^R(\omega=0)$.  Thanks to the relation (\ref{derivDelta}), the expression (\ref{lowfreqapprox}) involves  no new ingredient beyond the real and the imaginary part of the potential introduced in the previous section.  \\

After this preparation, we can now calculate the influence functional in the low frequency approximation. To do so, we use Eq.~(\ref{lowfreqapprox}), the relations $\Delta_{11}(0)=\Delta^R(0)+\iu\Delta^<(0)$, $\Delta_{12}(0)=\iu\Delta^<(0)$, $\Delta_{21}(0)=\iu\Delta^>(0)=\iu\Delta^<(0)$, and $\Delta_{22}(0)=-\Delta^A(0)+ \iu\Delta^<(0=-\Delta^R(0)+\iu\Delta^<(0)$, together with the definitions (\ref{eq:repot}) and (\ref{eq:impot}). We write the influence functional as  $\Phi[\Q] = \Phi_{_{QQ}}[\Q]+\Phi_{_{\bar Q\bar Q}}[\Q]+\Phi_{_{Q \bar Q}}[\Q]$. A straightforward calculation then yields
\be\label{QQ}
\Phi_{_{QQ}}[\Q] &=&
\frac{g^2}{2} \sum_{i,j=1}^N \int_{t_i}^{t_f} \diff{t}
\left[\frac{\!}{\!}V(\q_{j,2}-\q_{i,2})-V(\q_{j,1}-\q_{i,1})\right.\nn\\
&&-\iu W(\q_{j,2}-\q_{i,2})-\iu W(\q_{j,1}-\q_{i,1})+2\,\iu\,W(\q_{j,1}-\q_{i,2})\nn\\
&&+\left. \frac{\beta}{2}(\dot{\q}_{i,2}+\dot{\q}_{j,1})\cdot \frac{\partial}{\partial \q_{i,2}} W(\q_{j,1}-\q_{i,2})\right].
\ee
and similarly for $\Phi_{_{\bar Q\bar Q}}[\Q]$ with the substitution $\{\q_i\}\rightarrow\{\bar\q_i\}$. The mixed fermion-antifermion contribution reads
\be\label{QbarQ}
\!\!\!\!\!\!\!\!\!\!\!\!\!\!&&\Phi_{_{_{Q \overline Q}}}[\Q] =-g^2\sum_{i,j=1}^N \int_{t_i}^{t_f} \diff{t}
\left[\frac{\!}{\!} V(\q_{j,2}-\bar\q_{i,2})-V(\q_{j,1}-\bar\q_{i,1})\right.\nn\\
&&-\iu W(\q_{j,2}-\bar\q_{i,2})-\iu W(\q_{j,1}-\bar\q_{i,1})+\iu W(\q_{j,1}-\overline\q_{i,2})+\iu W(\q_{j,2}-\bar\q_{i,1})\nn\\
&&+\left. \frac{\beta}{2}\left(\dot{\bar\q}_{i,2}\cdot \frac{\partial}{\partial \bar\q_{i,2}}W(\q_{j,1}-\bar\q_{i,2})-\dot{\bar\q}_{i,1}\cdot \frac{\partial}{\partial \bar\q_{i,1}}W(\q_{j,2}-\bar\q_{i,1})\right)~\right].
\ee

Note that in the infinite mass limit, we can identify the coordinates of the heavy quarks in the upper branch of the contour with the corresponding ones in the lower branch, e.g.,  $\q_{j,1}=\q_{j,2}$. Furthermore, in this limit, we can ignore the velocity $\dot\q$. it is then easily verified that in this situation the influence functional vanishes identically. And this is as it should. We have then $P(\Q_f, t_f|\Q_i, t_i)=\delta(\Q_f-\Q_i)$. (See also the discussion at the end of the previous section.)

\subsection{Generalized Langevin equation}\label{sec:genLang}

We shall now use the expressions that we have obtained for the influence functional $\Phi$ in the low frequency approximation  in order to perform a further approximation that will lead us to a reformulation  in terms of a generalized Langevin equation. This last approximation exploits the fact that the trajectories of a heavy particle in the amplitude do not differ much from that in the complex conjugate amplitude. This suggests to perform the following change of variables 
\be\label{substitution}
\rr_i = \frac{1}{2}\left(\q_{i_1} + \q_{i_2}\right),\qquad 
 \y_i = \q_{i_1} - \q_{i_2},
\ee
(and similarly for the antiquarks coordinates), and to  expand the influence functional  in powers of the small deviations $\y_i$ and $\bar\y_i$.
In order to motivate this expansion, we note that, after an integration by parts,  the exponential of the free action takes the form
\be\label{eqn:largeM}
\exp\left[-\iu~M\sum_{i=1}^N \int_{t_i}^{t_f} \diff{t}(\ddot \rr_i\cdot\y_i)\right]\,,
\ee
and similarly for the antiquarks. The dominant contribution to the path integral comes from the region where the phase in Eq.~(\ref{eqn:largeM}) is small or at most of order unity. We can estimate the integral as  $\int_{t_i}^{t_f} \diff{t}(\ddot \rr_i\cdot\y_i)\sim \sqrt{T/M}|\y_i|$,  where $\sqrt{T/M}$ is the thermal velocity of the particle. The condition that the phase be small is then that $|\y_i|$ be small,  $|\y_i|\lesssim 1/\sqrt{MT}$.

We then proceed to the expansion of the influence functional to second order in $\y_i\,$. The details of this expansion are given in Appendix~\ref{expand}, and we report here the result. We collect the coordinates $\{\rr_i, \bar \rr_i, \y_i,\bar \y_i\}$ into $2N$ dimensional vectors\footnote{In fact these vectors have $2 N\times 3$ components since for instance $\rr_i$ is a three component vector. We do not indicate explicitly these components in order to alleviate the notation. Similarly, for each pair of vectors labelled by $i$ and $j$, say $\rr_i$ and $\rr_j$,  $\h_{\alpha\beta}(\rr_i-\rr_j)$ is a $3\times 3$ matrix mixing the components of the corresponding vectors.}, as we did earlier for $\Q$, e.g. $\Rg=(\rr_1,\cdots\rr_N,\bar\rr_1\cdots \bar\rr_N)$. As a result of the expansion, we can write the probability $P(\Rg_f,t_f|\Rg_i,t_i)$ as follows 
\be\label{langpath}
P(\Rg_f,t_f|\Rg_i,t_i) =\int_{\Rg_i}^{\Rg_f}\D\Rg\int_{\Yg_i=\zg}^{\Yg_f=\zg} \D\Yg~\exp\left[\int_{t_i}^{t_f}\rmd t\,{\cal L}(\Rg,\Yg)~\right],
\ee
where
\be\label{L}
{\cal L}(\Rg,\Yg)=\left( -\iu\,\Yg\cdot\left(M\ddot{\Rg}+M \bmgamma(\Rg)\cdot \dot{\Rg}-\mathbf{F}(\Rg)\right)-\frac{1}{2}~\Yg\cdot\bmlambda(\Rg)\cdot\Yg\right).\ee
We have $\Yg_i=\zg=\Yg_f$ because the coordinates $\q_{i,1}$ and $\q_{i,2}$ of the heavy particles coincide at the ends of the Schwinger-Keldysh contour.

The $2N$-dimensional vector $\mathbf{F}(\Rg)$ represents the forces between the heavy particles. It is given in terms of the gradient of the potential $V(\rr)$ as follows
\be\label{Force}
\mathbf{F}_{i^\prime}(\Rg)\equiv -g^2 \sum_{j=1}^N
\left(\begin{array}{rl}
\nab V(\rr_i-\rr_j) -\nab V(\rr_i-\bar\rr_j)\\
\\
\nab V(\bar\rr_{i}-\overline\rr_j)-\nab V(\bar\rr_{i}-\rr_j)
\end{array}\right)
\ee
where  $i=1,\dots,N$, and the primed index $i^\prime$ runs from 1 to $2N$, with $i=i^\prime$ for $i^\prime\le N$ ( first line of (\ref{Force})), $i=i^\prime-N$ for $i^\prime >N$ (second line of (\ref{Force})). The first line of Eq.~(\ref{Force}) represents the force exerted by all the heavy quarks and antiquarks on the $i^{th}$ heavy quark at position $\rr_i$, whereas the second line is the corresponding force exerted on the $i^{th}$ heavy antiquark at position $\overline\rr_i$.

The $(2N\times 2N)$-dimensional matrix $\bmgamma(\Rg)$ represents the friction exerted by the medium on the heavy particles. Its expression involves  the Hessian matrix $\h$ of the function $W$, the imaginary part of the potential, and reads 
\be\label{hessian}
\bmgamma_{i^\prime j^\prime}(\Rg)\equiv \frac{g^2}{2MT}\left(
\begin{array}{rl}
\h(\rr_i-\rr_j) & -\h(\rr_i-\bar\rr_j) \\
\\
- \h(\bar\rr_i-\rr_j) &  \h(\bar\rr_i-\bar\rr_j)
\end{array}\right),\qquad
\h_{\alpha\beta}(\rr)\equiv\frac{\partial W(\rr)}{\partial r_\alpha\partial r_\beta},
\ee
where the primed indices $i^\prime,j^\prime=1,\dots,2N\,$ are related to the unprimed ones, respectively $i$ and $j$, as indicated above. The Greek indices $\alpha, \beta,\gamma$ label the cartesian coordinates of $\rr$. The matrix  $\bmgamma$ is symmetric and real (hence diagonalizable  with real eigenvalues\footnote{ We shall see  that the eigenvalues are also strictly positive, which is physically expected for a matrix representing a friction term.}). This follows from the fact that, for instance, $ \h(\rr_i-\bar\rr_j) = \h(\bar\rr_j-\rr_i) $, and the fact that the $3\times 3$ matrix $\h_{\alpha\beta}(\rr)$,  being a Hessian matrix, is symmetric.
\\
Finally, the matrices $\bmgamma$ and $\bmlambda$ in Eq.~(\ref{L}) obey Einstein's relation
\beq
\bmlambda(\Rg)=2MT\bmgamma(\Rg).
\eeq

In the Appendix~\ref{appendixlangevin} we show that the probability (\ref{langpath}) can be generated by the following generalized Langevin equation \cite{Coffey}
\be\label{finallangev}
M\,\ddot{\Rg} = -M\bmgamma(\Rg)\cdot\dot{\Rg} + \mathbf{F}(\Rg) + \bm{\xi}(\Rg,t)\,,
\ee
with a   space dependent (also referred to as multiplicative) white noise $\bm{\xi}(\Rg,t)\,$:
\be\label{finaldev}
\langle\,{\xi_{i^\prime}}(\Rg,t)\,\rangle=0,\qquad
\langle\,{\xi}_{k^\prime}(\Rg,t)\,{\xi}_{m^\prime}(\Rg,t')\,\rangle = \bmlambda_{k^\prime m^\prime}(\Rg)\,\delta(t-t')\,.
\ee
The fact that the friction (and hence the noise) depends explicitly on the configuration of the heavy quarks is what makes this Langevin equation distinct from  what has been done so far in the context of heavy quark dynamics. The mathematical subtleties of such Langevin equations with multiplicative noise are recalled in the Appendix~\ref{appendixlangevin}. Let us just mention here that the present equation, with its explicit inertia term, does not suffer from discretization ambiguities, and we have used the Ito prescription to solve it numerically (see Appendices \ref{appendixlangevin} and \ref{algorithm} for details).

In order to get a first orientation as to the effect of the spatial dependence of the noise, we consider in the next subsection the simple case of a single pair of heavy particles, one heavy quark and one heavy antiquark, for which analytical results can easily be obtained. This will also be used to introduce the notion of  a bound state in this classical setting, and how such a bound state evolves when it is in contact with a thermal bath at various temperatures.

\subsection{Langevin equation for a single heavy quark antiquark pair}\label{langevinonepair}
When a  single heavy quark  antiquark pair is present in the system,  the generalized Langevin equation (\ref{finallangev}) takes the form
\be\label{finallang2}
 &&M~\ddot{\rr} + \frac{\beta\,g^2}{2}~\left(\h(0)\,\dot{\rr} - \h(\bms)\,\dot{\bar\rr}\right) - g^2~\nab V(\bms)  = {\xi}(\bms,t) \nn\\
&& M~\ddot{\bar\rr} + \frac{\beta\,g^2}{2}~\left(\h(0)\,\dot{\bar\rr}-\h(\bms)\,\dot{\rr}\right) + g^2~\nab V(\bms)  = {\bar\xi}(\bms,t)
\ee
where we have set $\bms\equiv \rr-\bar\rr$, with $\rr$ and $\bar\rr $ denoting the coordinates of the quark and the antiquark, respectively,
and the correlators of the noise are given by\footnote{Recall that $\h$ is a $3\times 3$ matrix, and that $\xi$ and $\bar \xi $ are here three dimensional vectors.}
\be
&&\langle~{\xi}_\alpha(\bms,t)~{\xi}_\beta(\bms,t')~\rangle = \langle~{\bar\xi}_\alpha(\bms,t)~{\bar\xi}_\beta(\bms,t')~\rangle = g^2~\h(0)\,\delta_{\alpha\beta}\delta(t-t')\nn\\
&&\langle~{\xi}_\alpha(\bms,t)~{\bar\xi}_\beta(\bms,t')~\rangle = -g^2~\h_{\alpha\beta}(\bms)\,\delta(t-t')\,.
\vspace{0.3cm}
\ee
The two Langevin equations are correlated through the force terms, as well as the friction terms which depend explicitly of the distance between the two heavy particles. It is convenient to write these equations in terms of relative ($\bms =\rr-\bar\rr$) and center of mass ($\bmrho=(\rr+\bar\rr)/2$) coordinates. By taking the sum and differences of the two equations above, we get
\be\label{finallang2b}
&& M~\ddot{\bmrho} + \frac{\beta\,g^2}{2}~\left[\h(0)- \h(\bms)\right]\dot{\bmrho} = \frac{{\bmxi}(\bms,t)+ \bar{\bmxi}(\bms,t)}{2}\nn\\
&& M~\ddot{\bms} + \frac{\beta\,g^2}{2}~\left[\h(0)+\h(\bms)\right] -2 g^2~\nab V(\bms)  = {\bar\bmxi}(\bms,t)-{\bmxi}(\bms,t).
\ee
Note that only $\bms$ is sensitive to the (attractive) force between the quark and the antiquark. The center of mass of the pair just follows a random walk and is subjected to a drag force and a random force. When the size of the pair exceeds the Debye radius, i.e., when $s \,m_D\gg 1$, $\h(\bms)\approx 0$, and the noises $\bmxi$ and $\bar\bmxi$ become uncorrelated. Using the fact that 
\beq\label{dragcoeff}
g^2~\h(0)_{\alpha\beta}= 2\,MT\,\gamma\,\delta_{\alpha\beta}
\eeq 
is diagonal, with $\gamma$ constant, we can then rewrite the equation for $\bmrho$ as a standard Langevin equation for a particle of mass $2M$ and drag force $\gamma$. In fact, in the limit $s m_D\gg 1$ the two equations (\ref{finallang2}) decouple. At large time, $|t-t_0|\gg \frac{1}{\gamma}$, the mean square displacement of $\rr(t)$ (and similarly for $\bar\rr(t)$) follows then the law of diffusion, 
\be\label{eq:brownian}
\langle\left(\rr(t)-\rr(t_0)\right)^2\rangle = 6\mathcal{D}\,|t-t_0|,\qquad \mathcal{D}=\frac{T}{M\,\gamma},
\ee 
where $\mathcal{D}$ is the diffusion coefficient.

In the opposite situation where $s m_D\ll 1$, the friction term cancels in the equation for $\bmrho$: this is because in this case the quark and the antiquark form an electric dipole of very small size that propagates in the plasma as a color neutral particle of mass $2M$, and hence does not interact with the plasma (one can easily verify that the contributions of the random forces also cancel, as they should). In the same limit of small size we can expand the potential, $V(\rr)\simeq V(0)+(1/2) k \rr^2$, with $k\equiv \left.\rmd^2 V/\rmd \rr^2\right|_{\rr=0}$ \footnote{We discuss in the next section how to regulate the Coulomb potential so as to give meaning to this expansion.}, and rewrite the equation for the relative motion as
\beq
\frac{M}{2}\ddot \bms+\frac{M}{2}\gamma\dot\bms- g^2 k\,\bms^2=\bmzeta(t),
\eeq
with $\langle~\zeta_i(t)~\rangle= 0$ and 
\be
\langle~\zeta_i(t)\,\zeta_j(t')~\rangle = \delta_{ij} \, MT\gamma\delta(t-t').
\ee
The heavy quark pair behaves then as a harmonic oscillator coupled to a thermal bath. This is as close as we can get to the notion of a bound state in this classical picture. Assuming that the expansion of the potential to quadratic order remains valid at large time, the mean square  displacement of $\bms(t)$ will eventually reach its value in thermal equilibrium, given by the equipartition theorem: 
\be\label{harmonicequilibrium}
\langle \bms^2\rangle= \frac{3\,T}{g^2k}.
\ee
This formula indicates that the radius of the $Q\bar Q$ pair increases with the temperature\footnote{We shall see in the next section that the  explicit linear dependence on $T$ of  the numerator is in fact amplified by the decrease of the coupling constant, and to a less extent that of the spring constant $k$, with increasing temperature}. As the temperature increases, the radius becomes eventually too large for the harmonic approximation to the potential to remain meaningful. In fact when this happens, the potential becomes essentially flat, indicating that no force maintains the  $Q\bar Q$ pair together: the bound state dissociates. These qualitative predictions will be made more quantitative in the next section.

\section{Numerical results}

We now present results of simulations of systems containing a given number $N$ of heavy quark antiquark pairs in a quark-gluon plasma at temperature  $T$. We discuss first the case of a single pair, $N=1$, and follow its fate for various temperatures, thereby turning the considerations of the previous subsection into a more quantitative  discussion. Then we turn to the case of many pairs (up to $N=50$), where, in addition to the phenomenon of dissociation that occurs for a single pair, the formation of new bound states through the process of recombination is also possible. \\

The  parameters in the problem are the mass $M$ of the heavy quarks, the temperature  $T$ of the plasma, and the gauge coupling $g$. In thermodynamical calculations, the latter quantity depends on the temperature and is commonly chosen to be the running coupling at a scale $\sim 2\pi T$. Although this is not crucial in the present work, we take into account this running of the coupling with  the following simple relation  taken from Ref.~\cite{Letessier}
\be\label{runningalpha}
\alpha_s = \frac{g^2}{4\,\pi}=\frac{\alpha_s(T_c)}{1+C\,\ln\left(\frac{T}{T_c}\right)},\quad C=0.760,\quad T_c=160 \mbox{ MeV},\quad \alpha_s(T_c)=0.5.
\ee
The Debye mass is approximated by its perturbative expression for a two flavor quark gluon plasma, $m_D^2=\frac{4}{3}\,g^2\,T^2$. With the running coupling given above we have $m_D\approx 460$ MeV for $T=T_c$.  The coupling of the heavy quark to the plasma constituents involve an extra color factor $C_F=4/3$ which is ignored. Finally we shall consider charm and bottom heavy quarks, whose masses are taken to be respectively $M_c=1.4$ GeV and $M_b=4.2$ GeV. Again, we emphasize that all these numbers, as well as all those which follow in this entire section,  are meant to provide reasonable orders of magnitude, in line with those expected for quarkonia in a quark-gluon plasma;  but we are not attempting to develop here a precise  phenomenology. 

In addition to the physical parameters that we have just discussed, we need to specify another one, a cutoff $\Lambda$, whose role is to control the short distance behavior of the real and imaginary parts of the heavy quark potential. This requires more discussion, and is the object of the next subsection. 

\subsection{Estimation of the cut-off}\label{Sec:cutoff}

Before we can use the Langevin equation derived in the previous section, we need indeed to cure two problems associated with the short distance behavior of the complex potential.
\begin{figure}
\begin{center}
\includegraphics[width=9cm]{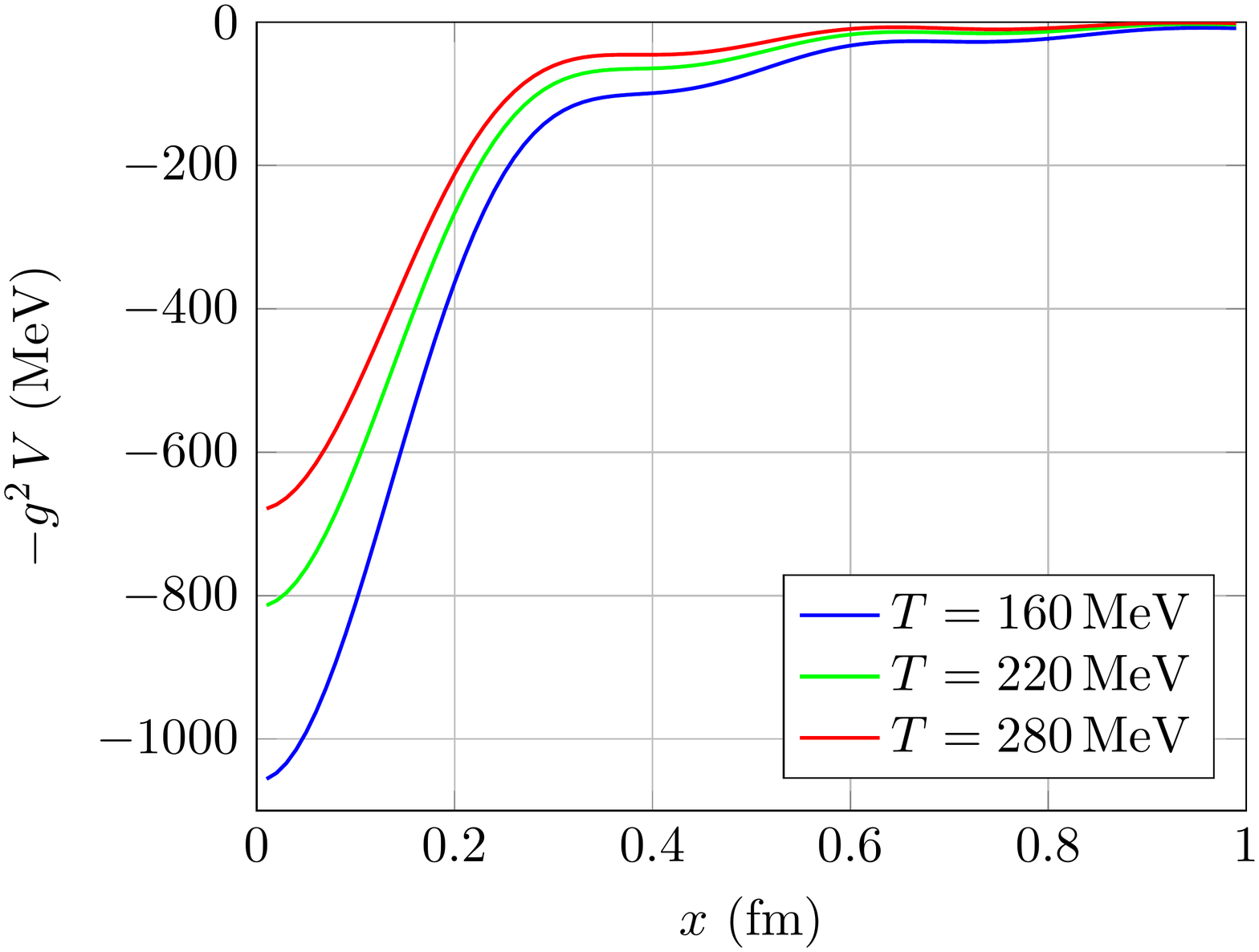}
\caption{The potential energy of a $Q\bar Q$ pair as a function of the $Q$-$\bar Q$ distance $x$ for three different temperatures,  and calculated with a cutoff $\Lambda=4$ GeV. Much of the temperature dependence  of the potential at small distance ($x\sim 0.05 fm$) can be attributed to that of the running coupling.   The temperature dependence of the screening mass $m_D\sim T$ affects the potential in the intermediate range ($x\sim 0.4 fm$). The oscillations at intermediate and large distances are an artifact of the finite cutoff. \label{potT}
}
\end{center}
\end{figure}

Consider first the real part  $V(\rr)$. This is given by the screened Coulomb potential, which behaves as $1/r$ at short distance. This poses a well known problem in classical simulations. One way to see it is to notice that the classical distribution,  that the Langevin equation eventually leads to, $\sim\rme^{-\beta V(\r)}$, is singular at small $r$ for the attractive Coulomb potential. This would lead to an infinite probability for two particles to be close together.  We may also  observe that when two particles come to close to each other, their relative kinetic energy becomes big, and this violates the conditions of validity of the approximations used in Sect.~\ref{low} when deriving the classical equations.  Note  that this is a problem that arises only in  the classical treatment of the Coulomb interaction through the Langevin equation; it would not occur if we were to solve the corresponding  Schr\"odinger equation. A simple way out is to add a repulsive ``quantum correction'' in the form  $\hbar^2/2Mr^2$, as originally proposed by  Kelbg \cite{Kelbg}. Many refinements of this procedure have been studied (in the present context, see for instance \cite{Dusling:2007cn} and references therein). In this exploratory work,  we find it sufficient to turn off the force at short distance, as was done for instance in \cite{Young:2008he}.  We do so here by introducing a finite cutoff in the integral  Eq.~(\ref{complexpotential}) that yields the screened Coulomb potential. The resulting potential is displayed in Fig.~\ref{potT}. Note that when calculated with this prescription the value of the potential at the origin, $V(0)$, depends linearly on the cutoff, $g^2V(0)\approx (2 \alpha_s/\pi)\Lambda$. Therefore the cutoff $\Lambda$ cannot be chosen too small otherwise the potential will not be deep enough to sustain bound states of the bottom quarks. Taking this into account, as well as further consideration to be presented shortly, we have settled for a value $\Lambda=4$GeV, and this is the value with which the plots in Fig.~\ref{potT} have been done. The temperature dependence  that is seen in Fig.~\ref{potT} arises mainly from the temperature dependence of the coupling constant, according to Eq.~(\ref{runningalpha}).

The presence of the cutoff makes the potential regular at short distance. One can then  expand it around the origin and find the spring constant $k$ introduced in Sect.~\ref{langevinonepair}. We get
\beq\label{kappa}
\frac{k}{m_D^3}=\frac{1}{6\pi^2}\left(\frac{\Lambda^3}{3 m_D^3}-\frac{\Lambda}{m_D}+ \arctan\frac{\Lambda}{m_D} \right).
\eeq
As mentioned in Sect.~\ref{langevinonepair}, the bound state will dissociate when the size, as measured by $\langle \bms^2\rangle=3T/(g^2k)$ becomes of the order of the Debye radius, $m_D^{-1}$. Defining the corresponding temperature as $T_D$,   we get (when $\Lambda\gg m_D$)
\beq
T_D\approx \frac{4\pi \alpha_s}{3} m_D \,\frac{k}{m_D^3}.
\eeq
For $\Lambda= 4$ GeV, $m_D=0.5$ Gev, this yields $T_D=320 $ MeV, a reasonable order of magnitude. 
This provides another argument in favor of a not too small cutoff.


\begin{figure}[htbp]
\begin{center}
\includegraphics[width=10cm]{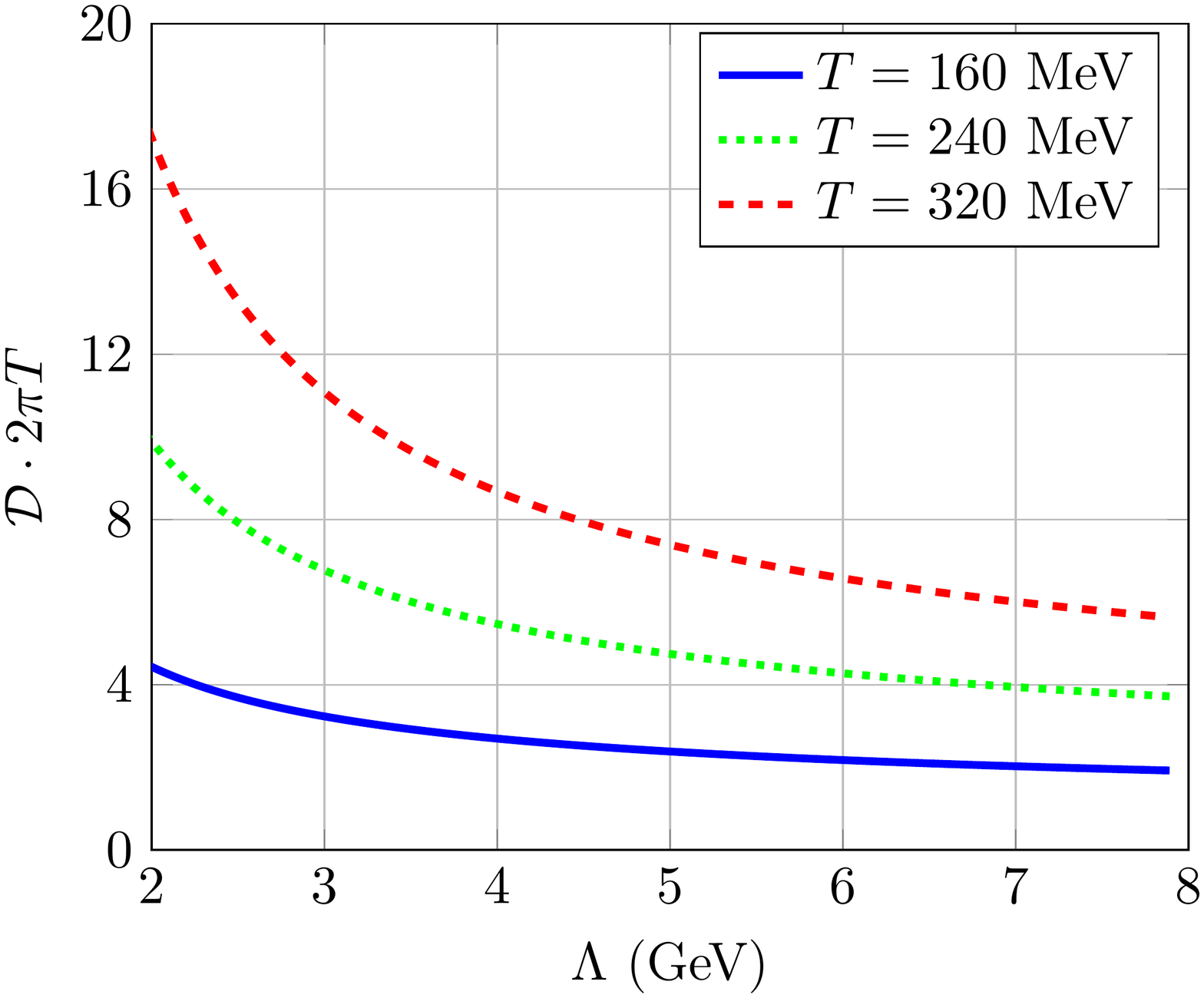}
\caption{The cutoff-dependence of the diffusion coefficient $\mathcal{D}=T/M\gamma$ (see Eq.~(\ref{eq:brownian})), multiplied by $2\pi T$, for different values of the temperature. Note that the differences between the three curves corresponding to different temperatures is largely due to the variation of the coupling constant, according to Eq.~(\ref{runningalpha}).}
\label{fig:D}
\end{center}
\end{figure}


The second reason why we need a cutoff is that the second derivative of the imaginary part of the potential, that enters the definition of the friction, is divergent. This was already mentioned at the end of Sect.~\ref{influence functional}. The problem here is of a different nature as that of the real part. It reflects the fact that the hard thermal loop approximation used in the calculation of the imaginary part of the potential involves kinematical approximations that cease to be valid whenever large momentum  exchanges are involved. Again the divergence can be controlled by a cutoff, which, here, would be naturally of the order of the temperature. In fact, we shall proceed as for the real part of the potential, and simply limit the momentum integral in Eq.~(\ref{eq:impot}) to values lower than $\Lambda$.  Note that  the values of $\Lambda$ that are needed for $V$ and $W$ are a priori unrelated to each other. However, for simplicity and in order to avoid the proliferation of irrelevant parameters, we have performed calculations with a common value for $\Lambda$, independent of the temperature. It turns out that  the drag coefficient and the diffusion constant  depend only mildly on $\Lambda$ around the value $\Lambda=4$ GeV that we have adopted (see Fig.~\ref{fig:D}).

From the second derivative of $W$ we can calculate the drag coefficient, according to Eq.~(\ref{dragcoeff})
and we get
\be\label{lambdarel}
\gamma=\frac{m_D^2\,e^2}{24\,\pi\,M}\left( \ln\left(1+\frac{\Lambda^2}{m_D^2}\right)-\frac{\Lambda^2}{\Lambda^2+m_D^2} \right),
\ee
To within a color factor $C_F$ that we ignored, and with the specific choice $\Lambda=T$, this expression agrees with that obtained in Ref.~\cite{Moore:2004tg} in the leading logarithm approximation. 
The diffusion constant $\mathcal{D} = {T}/({M\,\gamma})$ is plotted in  Fig.~\ref{fig:D} as the dimensionless combination $(2\pi T) {\cal D}$:
\beq
\mathcal{D}\cdot 2\pi T=\frac{9}{4\alpha_s^2}\left(\ln\left(1+\frac{\Lambda^2}{m_D^2}\right) - \frac{\frac{\Lambda^2}{m_D^2}}{\frac{\Lambda^2}{m_D^2}+1}\right)^{-1}.
\eeq
One sees that in the region  $\Lambda\simeq 4$ GeV, the diffusion constant depends indeed weakly on the value of $\Lambda$. Furthermore, for this value, $2\pi T{\cal D}\approx 2.7$ for $T=160$ MeV, or $\gamma\approx 0.2$ fm$^{-1}$. These values are of the order of magnitudes of those used in phenomenological studies \cite{Moore:2004tg} (see also \cite{Das:2015ana} for more recent estimates). \\

Now that we have adjusted all the parameters, we can start exploring the main features of the dynamics of the heavy quarks in a plasma, as predicted by the generalized Langevin equation (\ref{finallangev}). The details of the numerical method that we use to solve this equation are given in Appendix~\ref{algorithm}.

\subsection{One heavy quark-antiquark pair}

Our first set of results concerns the evolution of a heavy $Q\bar Q$ pair immersed in  a uniform quark-gluon plasma in thermal equilibrium at temperature $T$. The pair is prepared so that it corresponds initially to a bound state with a given size and binding energy.  One first generates a sample of pairs, with the following procedure: The distance between the quark and the antiquark is chosen randomly between $0$ and the Debye radius $r_D=m_D^{-1}$. The relative initial velocity of the quark and the antiquark is taken from a Maxwell distribution centered at the average value of typical quarkonia relative velocities (e.g. $v_0^2\sim 0.3$ for charmonium \cite{bodwin}).   Then we select from this sample the pairs that can be associated with specific bound states according to criteria that will be specified shortly. 
We then simulate the evolution of the pair using the Langevin equation (\ref{finallangev}) that was derived in Sect.~\ref{langevinonepair}.

\begin{figure}[htbp]
\begin{center}
\includegraphics[width=14cm]{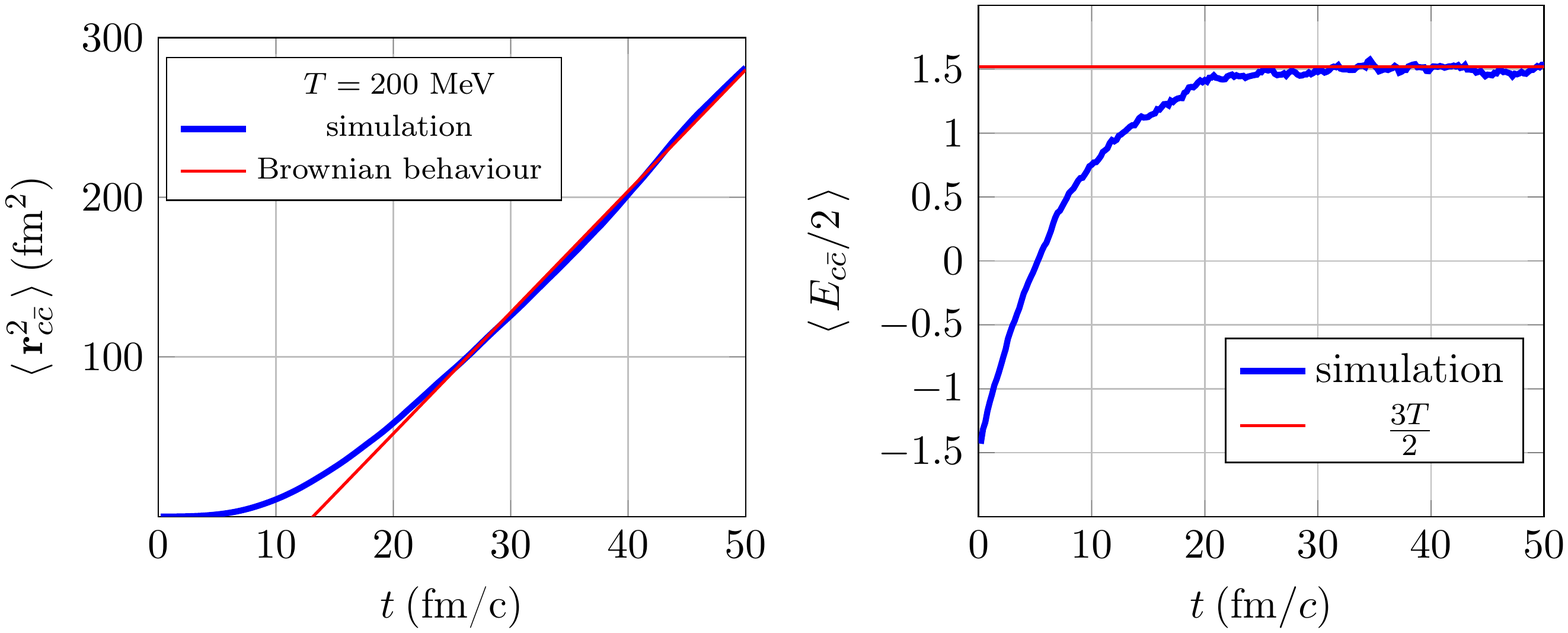}
\caption{On the left: Average  $c$-$\bar{c}$ distance squared as a function of time. This follows the predicted long-time Brownian behavior  with the diffusion constant given by $\mathcal{D}\cdot2\pi T\approx 4$. On the right: Average energy (in $200$ MeV units) per quark as a function of time compared with the energy at equilibrium (horizontal line). In both graphs we used $T=200$ MeV and $\Lambda=4$ GeV. Statistical errors are to small to be plotted.}
\label{fig:brown}
\end{center}
\end{figure}

As a first check of the Langevin dynamics, we consider a  $c\bar{c}$ pair at a temperature $T=200$ MeV. At this temperature, and for the parameters that we have chosen, all $c\bar c$ bound states eventually dissociate in the plasma. This is what the plot on the left of Fig.~\ref{fig:brown} indeed shows. After an initial transient period of time, the two consituents of the pair follow independent  Brownian motions, with the average  distance squared growing linearly with time, in agreement with the analytical result, Eq.~(\ref{eq:brownian}). Moreover,  the right panel of Fig.~\ref{fig:brown}  shows that the constituents indeed thermalize, the energy per quark reaching the value $(3/2)T$, in agreement with the equipartition theorem.

However, the very long time,  where the heavy quarks eventually thermalize with the surrounding plasma, is not our main concern here. We want to understand the dynamics over shorter time scales, in particular because the  plasma produced in a nucleus-nucleus collisions has a finite lifetime. To be specific, we shall take  this lifetime to be  $\tau_{\rm qgp}\sim10\,$fm/c, and accordingly our main focus will be to understand the dynamics of the heavy quarks over such a typical time scale. We shall also differentiate between different charmonium states, $J/\Psi$ ($\,1$S), $\chi_c$ ($\,1$P) and $\Psi'$ ($\,2$S), but consider a single  bottomonium state which we shall refer to as  the $\Upsilon$. A word of explanation is needed here regarding what we mean by \textit{bound states}.  Within the classical simulation using the Langevin equation, this refers to the following procedure. At the beginning of the simulation we calculate the binding energy of a pair in its center of mass frame\footnote{The binding energy is known at each time step of the simulation, since we follow both the velocities and the positions of the particles.} and select the pairs according to the values of their initial radius $r_0$, and their binding energy $\Delta E\,$. Depending on these values, we call a pair by the name of the closest bound state it would correspond to in a complete quantum treatment. The specific criteria that we use to attribute a charmonium state to a given pair are the following
\begin{itemize}
\item $\Psi'\qquad$ :  $  0<\Delta E< 100$ MeV and $r_0\geq 0.35$ fm,
\item $\chi_c\qquad$ : $100\leq\Delta E\leq 300$ MeV and $r_0\geq 0.25$ fm, 
\item $J/\Psi\quad\;$ : $\Delta E\geq 550$ MeV and $r_0\geq 0.10$ fm,
\item $\Upsilon\qquad$ : $\Delta E\geq 700$ MeV.
\end{itemize}
For the  bottomonium, as already mentioned, we do not attempt to discriminate between the various bound states, and the requirement of a large binding energy automatically selects  small sizes. In the case of  charmonia, the constraint on the radius discriminates form instance a $c\bar c$ pair with the binding energy of a $\Psi'$ but  the radius of a $\chi_c$,  and so forth. For the $J/\Psi$ the minimum radius $r_0=0.1$ fm  eliminates too high values of the binding energy. Such requirements do not apply to the $\Upsilon\,$,  the binding energy being in this case limited by the depth of the potential (controlled by the cutoff $\Lambda$, as discussed in the previous subsection).

\begin{figure}
\begin{center}
\includegraphics[width=15cm]{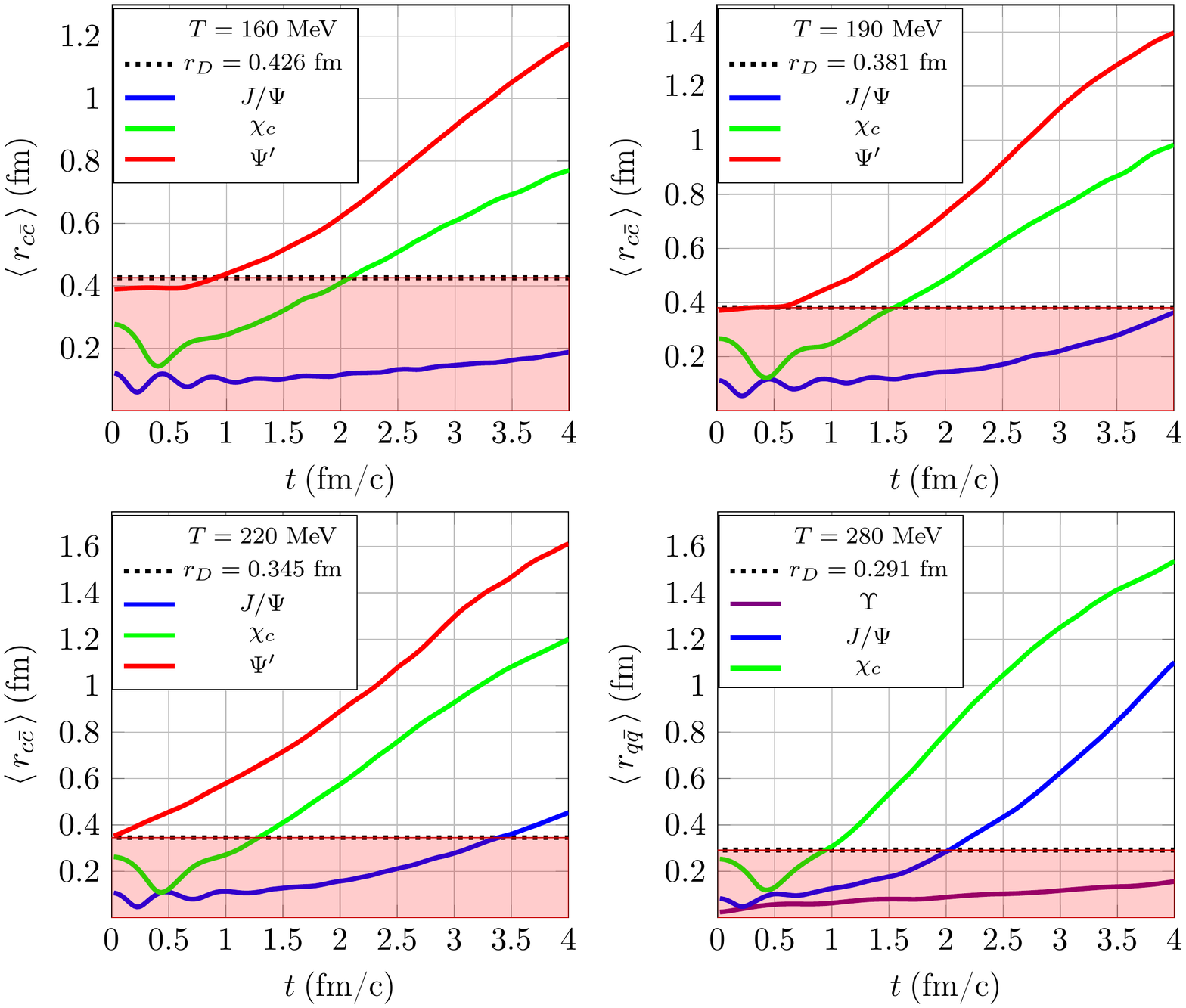}
\caption{(Color online.) Average quarkonia radius  as a function of time. The pairs are prepared as ``bound states'' following the procedure explained in the text. The shaded part indicates the region in which the charmonia radii are smaller than the Debye radius $r_D$. Statistical errors are too small to be plotted. In each plot, the upper curve represent the less bound system, the lower curve the most bound one.}
\label{fig:brown-plateau}
\end{center}
\end{figure}

The time evolutions of the average size of pairs thus prepared are presented in 
Fig.~\ref{fig:brown-plateau} for different temperatures. The harmonic oscillator pattern expected from the analysis of Sect.~\ref{langevinonepair} for pairs of small initial sizes is clearly visible.  There are indeed cases where $\langle\,r_{\rm q\bar q}\,\rangle$ remains almost constant for a certain time interval, reflecting the fact that the corresponding pair is highly correlated, or ``bound''. The lower  the temperature, the longer the correlation lasts. One also observes the expected ``sequential melting''  of $\Psi'\,$, $\chi_c\,$, $J/\Psi$ and $\Upsilon$ as temperature grows. Of course  the sequential dissociation of $\Psi'\,$, $\chi_c\,$, $J/\Psi$ just reflects the inequalities of their respective sizes,  $r_{\Psi'}>r_{\chi_c}>r_{J/\Psi}\,$. One may try and attribute different ``melting temperatures'' to the dissociation of the various bound states. 
For example, we see from Fig.~\ref{fig:brown-plateau} that the initial plateau associated with the average $\Psi'$ radius is absent at $T=220$ MeV, indicating that $\Psi'$ immediately dissolves at this temperature, while the plateau is still visible at $190$ MeV. One may then infer  that the melting temperature of $\Psi'$ is $T\approx 200$ MeV. 
Using the same argument of the size of the screening radius, we can extract a melting temperature of $T\approx 310$ MeV for $\chi_c\,$.
On the other hand, it is evident that the $J/\Psi$ survives up to much higher temperatures than the other two charmonium states. However, for the $J/\Psi$ we can not use the above strategy to estimate its melting temperature, because of the limitation of the numerical setup:  when the temperature increases ($\,T\gtrsim 400$ MeV) it becomes impossible (with the present choice of parameters) to prepare an initial $J/\Psi$ with $\Delta E\geq 550$ MeV, the potential well is simply not deep enough (see Fig.~\ref{potT}). Later, we shall estimate the melting temperature of the $J/\Psi$ by using a different procedure.

In the last panel of Fig.~\ref{fig:brown-plateau} we also compare the $\chi_c\,$, $J/\Psi$ and $\Upsilon$ behaviours at $T=280$ MeV. We see that the average $b\bar{b}$ pair is far more strongly correlated than the $c\bar{c}$ pair, and the $\Upsilon$ radius remains small ($\,\langle\,r_{\Upsilon}\,\rangle\leq r_D\,)$ for a relatively long time (we shall see in the next subsection that the melting temperature of the $\Upsilon$ ($\,1$S) state is $T>600$ MeV).

\begin{figure}[htbp]
\begin{center}
\includegraphics[width=16cm]{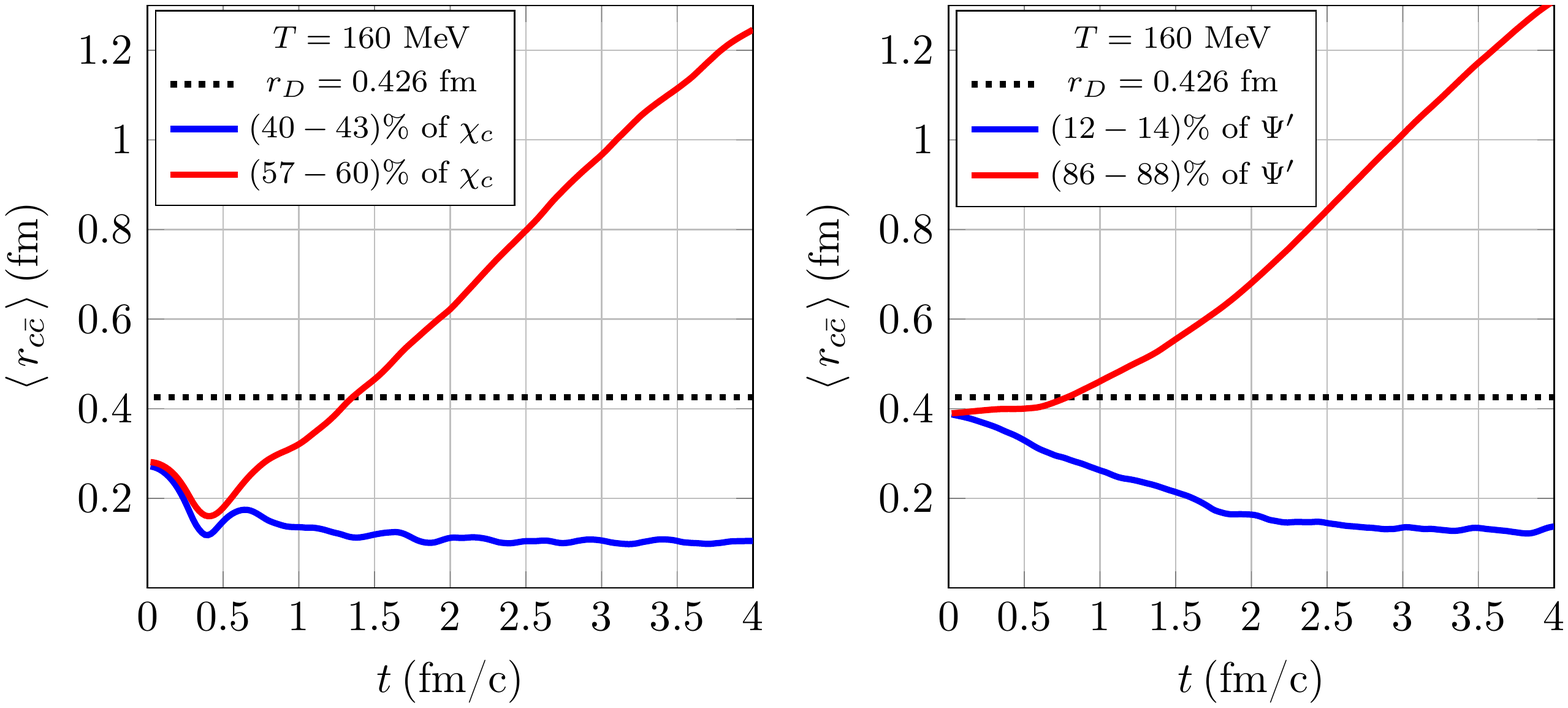}
\caption{(Color online.) On the left: Comparison between $\langle\,r_{c\bar c}\,\rangle$ of $\chi_c$ states that become $J/\Psi$ ($\approx 41\%$ -- lower curve) and those that do not decay ($\approx 59\%$ -- upper curve). On the right: Same comparison for $\Psi'\,$. The pairs are prepared as bound states as indicated in the text. For similar initial conditions, they evolve statistically to different final states. }
\label{fig:pure}
\end{center}
\end{figure}

The curves representing the $\chi_c$ exhibit an interesting phenomenon. One sees that in all cases, the corresponding radius tends to decrease initially, bringing the $\chi_c$ closer to a more stable bound state ($J/\Psi$). This is a clear indication that, at these temperatures, the binding forces are not yet entirely screened. While on average, the relative kinetic energy prevents the $\chi_c$ to really decay into a $J/\Psi$, as the curves in Fig.~\ref{fig:brown-plateau} indicate,  a substantial fraction of the pairs prepared as $\chi_c$ does decay into $J/\psi$'s, as shown in Fig.~\ref{fig:pure}. 
It is possible to estimate the percentages of $\chi_c$ and $\Psi'$ states that decay into $J/\Psi\,$, a process  known as \textit{feed-down}. In Table~\ref{feed-down} the feed-down percentages of $\chi_c$ and $\Psi'$ are listed for some values of the temperature. We found that, for each temperature, there are more $\chi_c$ than $\Psi'$ states that decay into $J/\Psi\,$, and the feed-down mechanism decreases when the temperature grows: the more fragile states prefer to dissociate rather than form a more strongly bound system.  Amusingly, the feed-down fractions   obtained here  at $T=190$ MeV are similar to the experimental values quoted in \cite{Faccioli}, although of course the physical context is rather different. 
\begin{table}
\begin{center}
\end{center}
\centering
\begin{tabular}{ccccc}
\toprule
$T$ (MeV)       & 160           & 190        & 220       &    280    \\
\midrule
 $\chi_c$       &  (40-43)\%    &  (28-30)\% & (16-17)\% &   (1-2)\%  \\
\\
 $\Psi'$        &  (12-14)\%    &  (7-8)\%   &   0\%     &     /      \\
\bottomrule
\end{tabular}
\caption{Fractions of $\chi_c$ and $\Psi'$ eventually becoming $J/\Psi$'s.}
\label{feed-down}
\end{table}
In Fig.~\ref{fig:pure} we compare the different behaviors of the $\chi_c$ and $\Psi'$ average radii, separating those which decay from those which do not. Looking on the left of Fig.~\ref{fig:pure} we notice that even the non-decaying $\chi_c$ states initially reduce (on average) their radius (also at $T=280$ MeV, as seen in the last panel of Fig.~\ref{fig:brown-plateau}). This why their average lifetime (see Table~\ref{tab:lifetime}) remains almost the same below the melting temperature ($\approx 310$ MeV), whereas the average lifetime of a non-decaying $\Psi'$ diminishes as the temperature goes up.

The lifetimes of Table~\ref{tab:lifetime} are calculated by averaging the time intervals needed for the radii of $\chi_c\,$, $\Psi'$ (both not-decaying) and $J/\Psi$ to become larger than the Debye screening length. One notices that the $J/\Psi$ lifetime at $T=280$ MeV is still quite appreciable. 

\begin{table}[htbp]
\begin{center}
\end{center}
\centering
\begin{tabular}{ccccc}
\toprule
$T$ (MeV)       & 160           & 190        & 220       &    280    \\
\midrule
 $J/\Psi$       &  $\gtrsim 10$    &  $\gtrsim 10$ & 4.9$\pm$ 0.2 &   2.8$\pm$ 0.2  \\
\\
 $\chi_c$       &   1.6$\pm$ 0.1   &  1.6$\pm$ 0.1 & 1.5$\pm$ 0.1 &   1.6$\pm$ 0.1  \\
\\
 $\Psi'$        &  0.7$\pm$ 0.1    &  0.5$\pm$ 0.1   &   0.1$\pm$ 0.1     &    0     \\
\bottomrule
\end{tabular}
\caption{Average charmonium lifetimes (in fm/c) in the quark-gluon plasma. Only the  $\chi_c$ and $\Psi'$ that do not decay into $J/\Psi$ are taken into account in the lifetime estimates.  }
\label{tab:lifetime}
\end{table}

\newpage

\subsection{Many heavy quark-antiquark pairs}
\vspace{0.5cm}

In the previous subsection, we saw how a single heavy quark antiquark pair can evolve from an apparent bound state to a system of two independent quarks that eventually thermalize with the plasma on long time scales. We could also observe, with a proper selection of the initial conditions the expected phenomenon of  sequential dissociations. Finally, we provided some criterion to get a crude estimate of the lifetime of the bound state in the plasma. We would like now to examine how these features are modified when several pairs are present in the plasma.
\begin{figure}[htbp]
\begin{center}
\includegraphics[width=10cm]{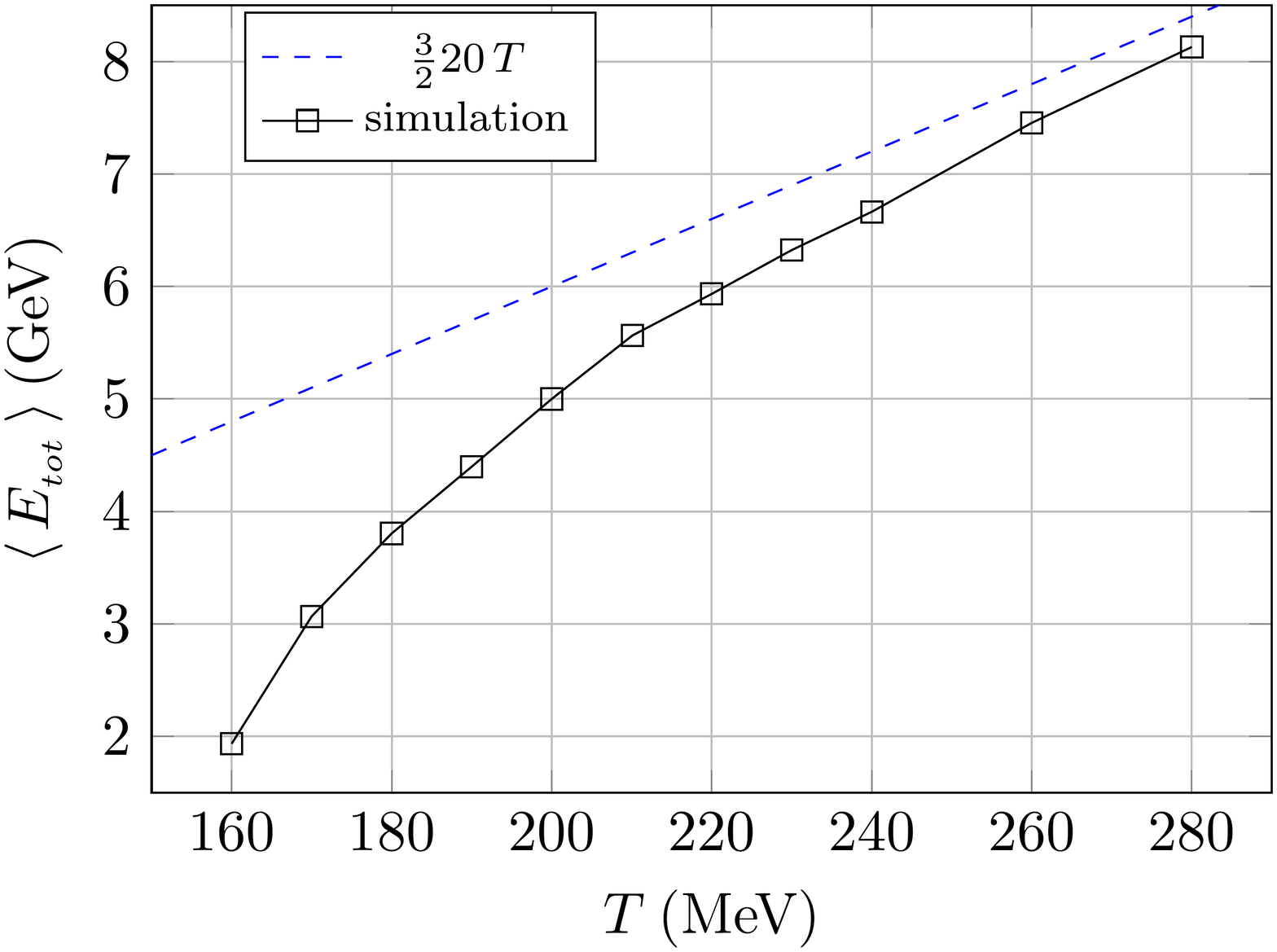}
\caption{Average total energy of a system of $10$ $c\bar c$ pairs in thermal equilibrium as a function of temperature. Before measuring the energies, we ran the simulations for a time interval of $100$ fm in order to let the system thermalize. Simulations were performed in a periodic cubic box of side $4$ fm and statistical errors are again too small to be plotted.}
\label{fig:cv}
\end{center}
\end{figure}
The simulations that we shall present were performed for $N=2, 10$ and $50$ quark-antiquark pairs, in a cubic box of side $4$ fm, with periodic boundary conditions. \\

When there are enough pairs in the system, one expects them to evolve towards an ideal gas of the constituents, if the temperature is high enough. The average energy for a system of $N=10$ pairs is plotted in Fig.~\ref{fig:cv}, and compared to that of an ideal monoatomic gas of $2N$ particles, 
\be
E_{\rm gas} = \frac{3}{2}(2N)\,K_BT. 
\ee
The expected trend is clearly visible, and at the largest temperatures considered, $T\gtrsim 280$ MeV, the ideal gas limit is almost reached. At such high temperatures, most of the pairs dissociate if one waits long enough. On the other hand,  at lower temperatures, pairs may survive and this results in  the average energy of the system being lower than that of the ideal gas  at the same temperature. We note that the process of dissociation, considered from this thermodynamical point of view, is a gradual process: even at high temperature there remains some finite probability to find a bound pair. Given the length of the simulation (over $100$ fm/c), and that, in this range of temperatures, a single pair would eventually dissociate, the equilibrium state that we are observing results from the balance of the two competing effects of dissociation and recombination, as we shall discuss in more details shortly. 

\begin{figure}[htbp]
\begin{center}
\includegraphics[width=10cm]{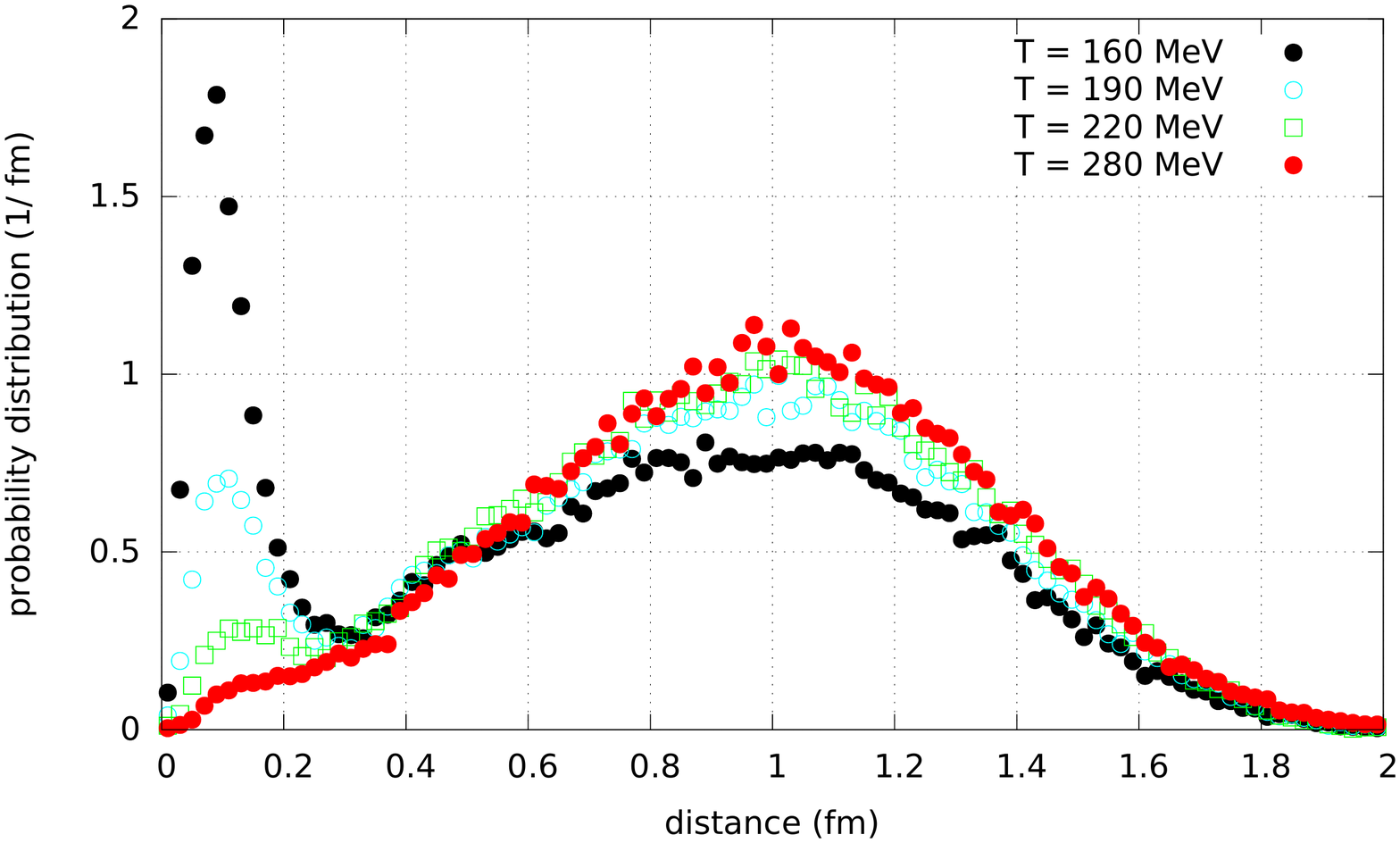}
\caption{The figure shows the distribution of distances to the nearest antiquark from a given quark. This probability is computed in the following way: once thermal equilibrium is reached, one computes from each quark  the distance to the nearest antiquark, draw an histogram, and normalize in order to get the distribution. Simulations were performed for a system of 10 $c\bar c$ pairs in a cubic box of side $4$ fm, with periodic boundary conditions.}
\label{fig:closest}
\end{center}
\end{figure}

The presence of bound pairs in the system can also be inferred form the analysis of another quantity that is directly sensitive to the correlations between two particles, namely the  probability distribution $P_{q\bar q}$ of the distance from a given quark to the nearest antiquark. In an ideal gas, this distribution is given by 
$$
P^{\rm ideal}_{q\bar q}(r) = \frac{3}{a}\!\left(\frac{r}{a}\right)^{\!2}\!\left(1-\left(\frac{r}{a}\right)^3\!\frac{1}{N}\right)^{\!\!N-1}\stackrel{N\gg 1}{\simeq} \frac{3}{a}\left(\frac{r}{a}\right)^2 e^{-(r/a)^{\frac{1}{3}}}\:,
$$
where $a=\left(\frac{3}{4\pi\rho}\right)^{1/3}$ is the mean distance between the antiquarks and $\rho=\frac{N}{V}\,$ the density of antiquarks. The peak of the ideal gas distribution for $N=10$  quark-antiquark pairs in a cubic box of side $4$ fm, is at $r_{\rm peak}=(\frac{20}{29})^{1/3}a \approx 1.15$ fm.
This peak is clearly visible in the distribution $P_{q\bar q}$ of the interacting system  which is plotted in Fig.~\ref{fig:closest}. But this figure reveals  also another feature: at low temperature, there is also a sharper peak reflecting the presence of highly correlated states in the system. These, we associate with the bound states. In line with the previous plot, Fig.~\ref{fig:cv}, this peak disappears when  $T\gtrsim 280$ MeV.
From Fig.~\ref{fig:closest} we can also infer that a correlated $c\bar{c}$ pair has a maximum radius of approximately $0.3$ fm, which is indeed similar to the values of the Debye screening length in this range of temperature.\\

\begin{figure}[htbp]
\begin{center}
\includegraphics[width=7.4cm]{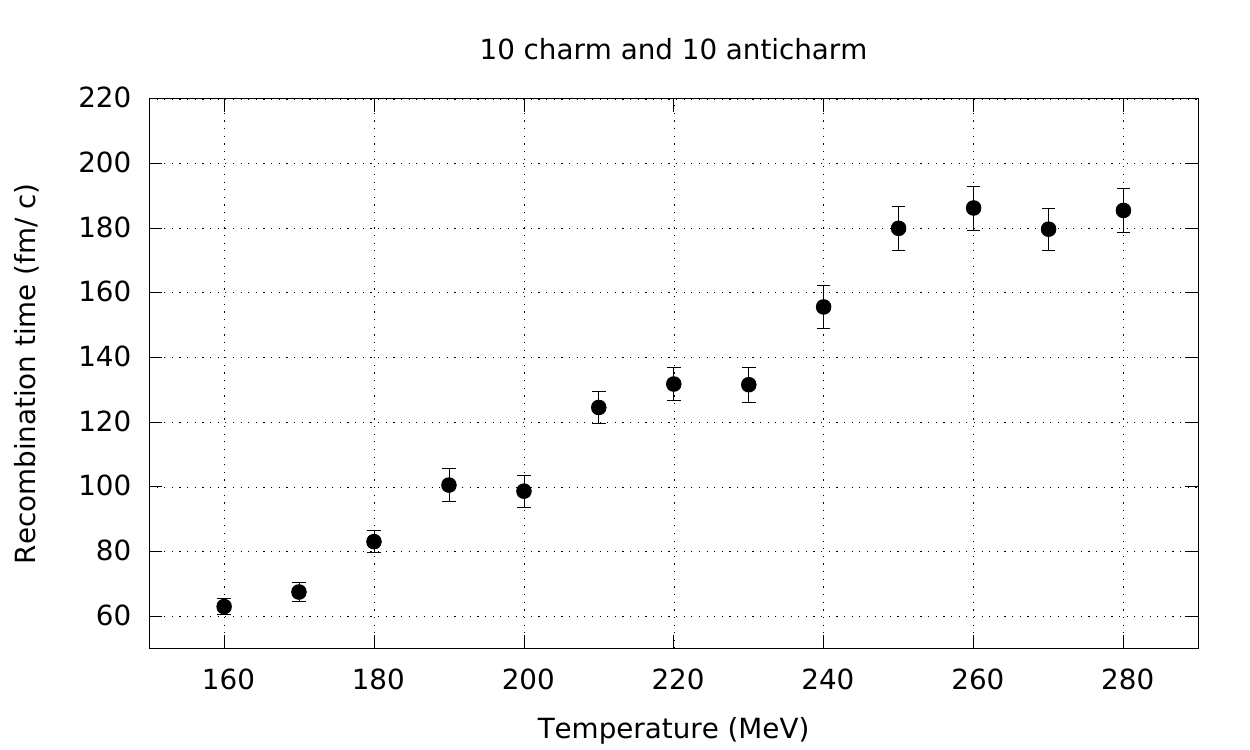}
\includegraphics[width=7.4cm]{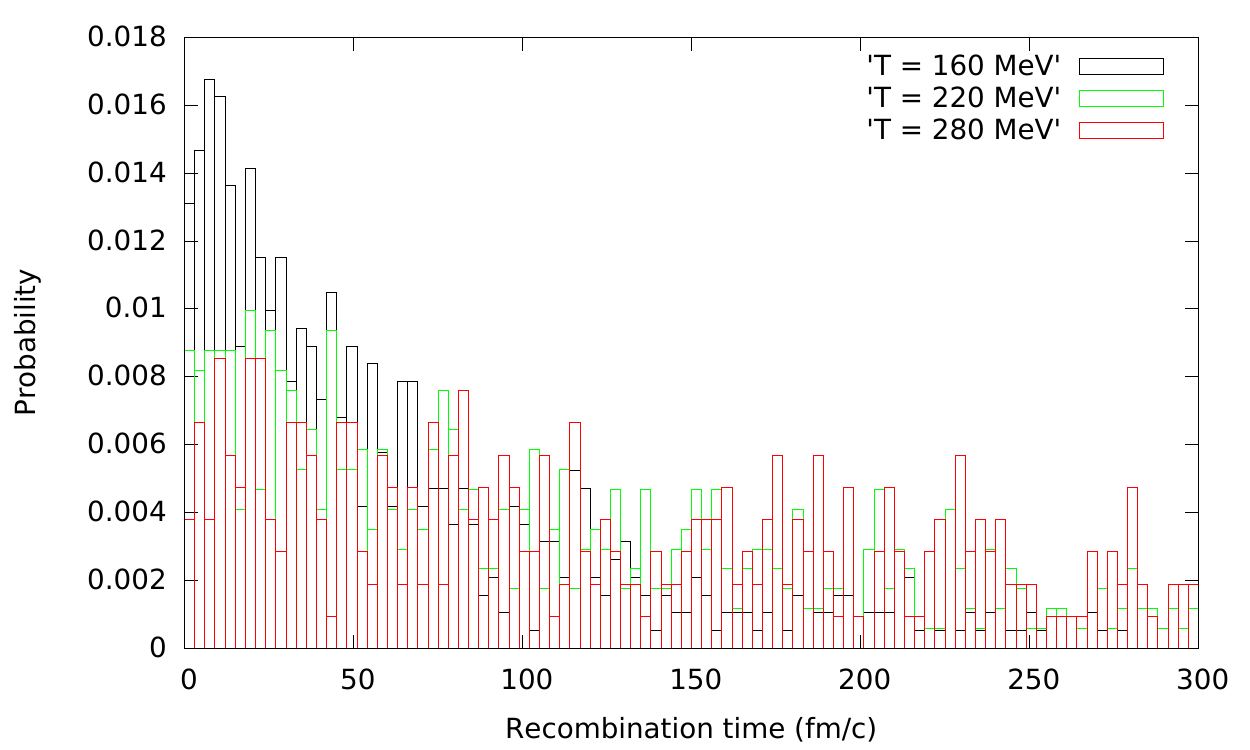}
\caption{On the left: Average recombination time as a function of temperature. On the right: Probability of recombination times for three values of temperatures. The quark-antiquark distance has been chosen to be less or equal to $0.3$ fm for the quark-antiquark configuration to be considered a pair (see text). Both simulations were performed with a system of 10 $c\bar c$ pairs in a cubic box of side $4$ fm, with periodic boundary conditions.}
\label{fig:rec} 
\end{center}
\end{figure}

We turn now to a more detailed study of the process of recombination.  We start with the evaluation of the recombination times for the pairs as a function of the temperature, that is the average time needed for a quark (antiquark) to form a pair, once the quark (antiquark) moves away from its previous antiquark (quark) partner.
In doing this calculation we carefully avoid counting  the contributions of ``non-interacting'' events, that is the occurrences where a quark passes by an antiquark without forming an actual pair.
In order to eliminate such events,  we performed simulations for  a non-interacting system with a constant (space-independent) drag constant (see Eq.~(\ref{dragcoeff})) and we calculated the corresponding normalized distribution $P_{\rm free}(t)$ of the time intervals $t$ in which a charm and an anticharm stay close together within a sphere of radius $0.3$ fm. Then, for each temperature, we define a minimum lifetime   $\tau$ by the condition
\be\label{eq:1percent}
\int_\tau^\infty \rmd t\,P_{\rm free}(t) < 1\%\:.
\ee
By selecting pairs that stay together for a time greater than $\tau$, only pairs formed because of the interactions (and not those resulting from random encounters) contribute to the recombination times. Note that the procedure does not allow for a  detailed analysis in terms of various bound states, as we were able to do for the dissociation: the small lifetimes typical of $\chi_c$ and $\Psi'$ are automatically discarded by the procedure, so that we implicitly consider only $J/\Psi$ regeneration.

As one can see on the left of Fig.~\ref{fig:rec}, the outcome for a system of $10$ $c\bar c$ pairs is that the recombination time increases almost linearly with the temperature, starting from a value of $t_{\rm rec} = (62.9 \pm 2.5)$ fm at $T=160$ MeV and reaching a value of $t_{\rm rec} = (185.5 \pm 6.8)$ fm at $T=280$ MeV.
As one increases the temperature one increases the rate of encounters, but also the relative kinetic energies of the pairs, preventing binding. 
Another important observation is that the recombination times are very long, so long that one may wonder whether the mechanism of recombination could be of any phenomenological relevance. However, as the  graph on the right panel of Fig.~\ref{fig:rec} shows, the distribution  of the recombination times is very broad. Thus, even if the standard errors of the graph on the left panel of Fig.~\ref{fig:rec} are small (because of the large statistics), the standard deviations are of the same order as the average values: over  the lifetime of the quark-gluon plasma ($\sim 10$ fm/c) there is effectively no characteristic time scale for recombination.

\begin{figure}[htbp]
\begin{center}
\includegraphics[width=7.5cm]{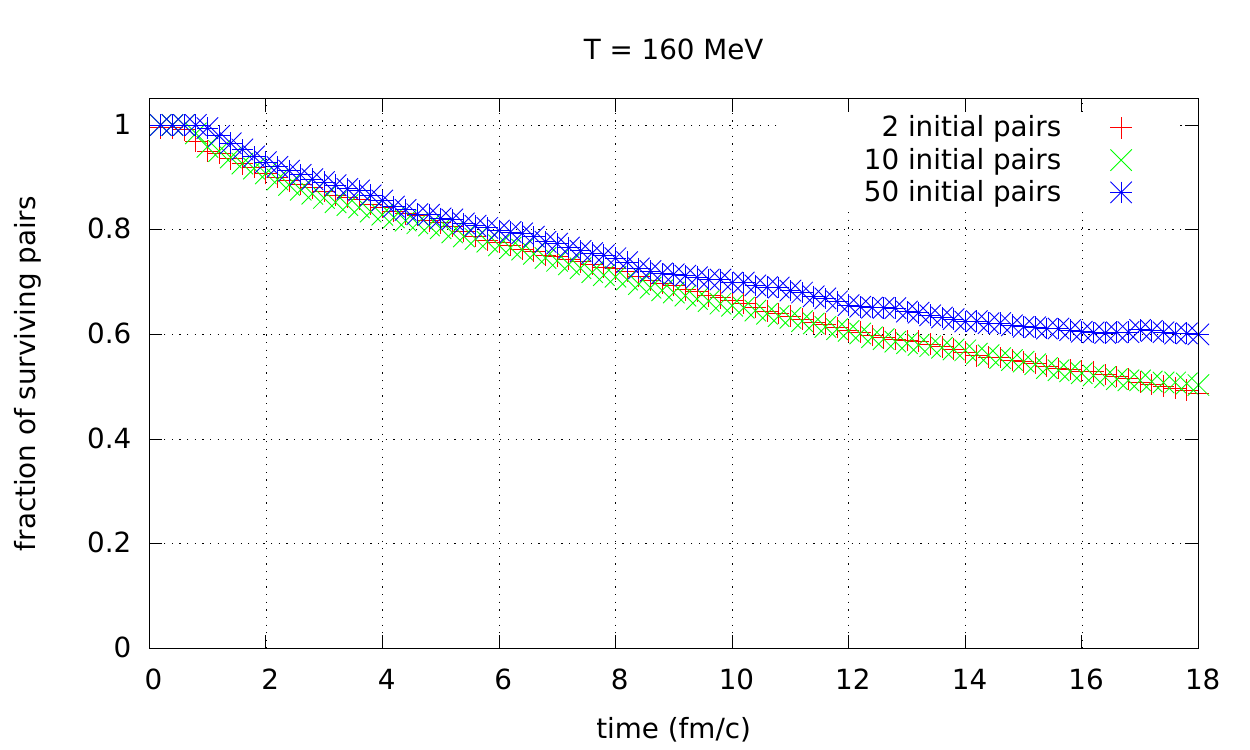}
\includegraphics[width=7.5cm]{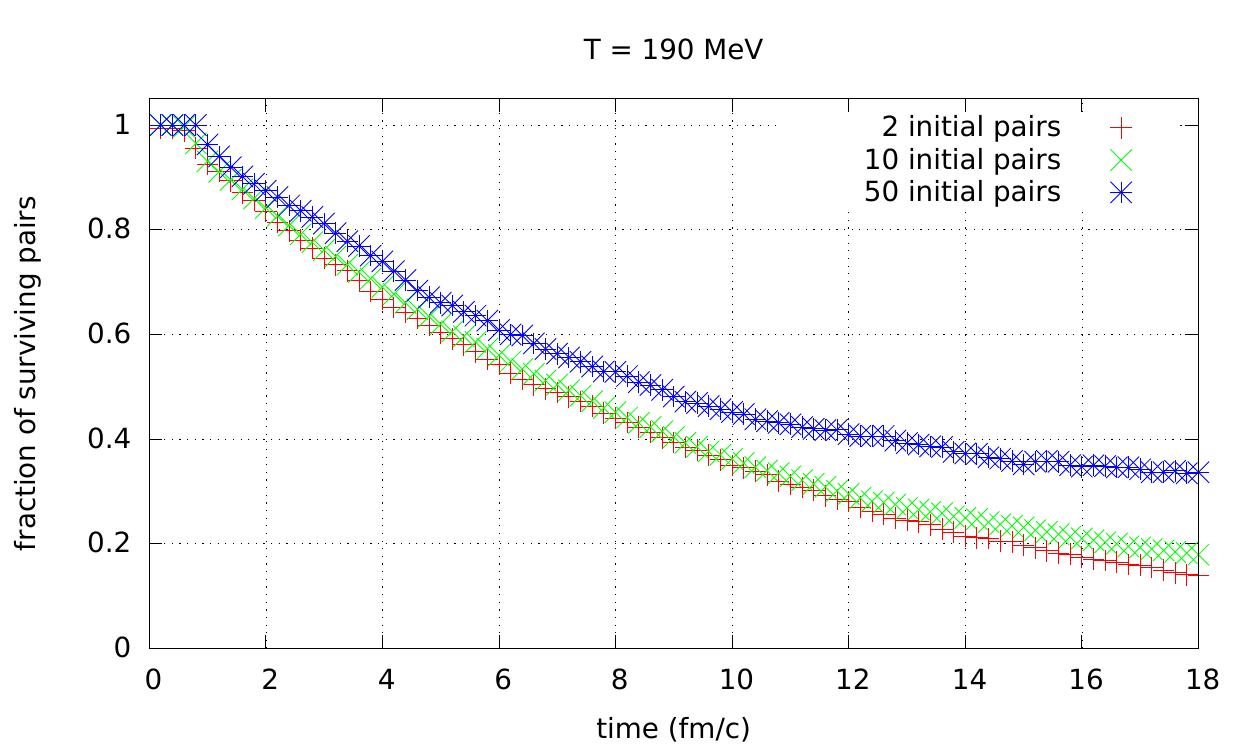}
\includegraphics[width=7.5cm]{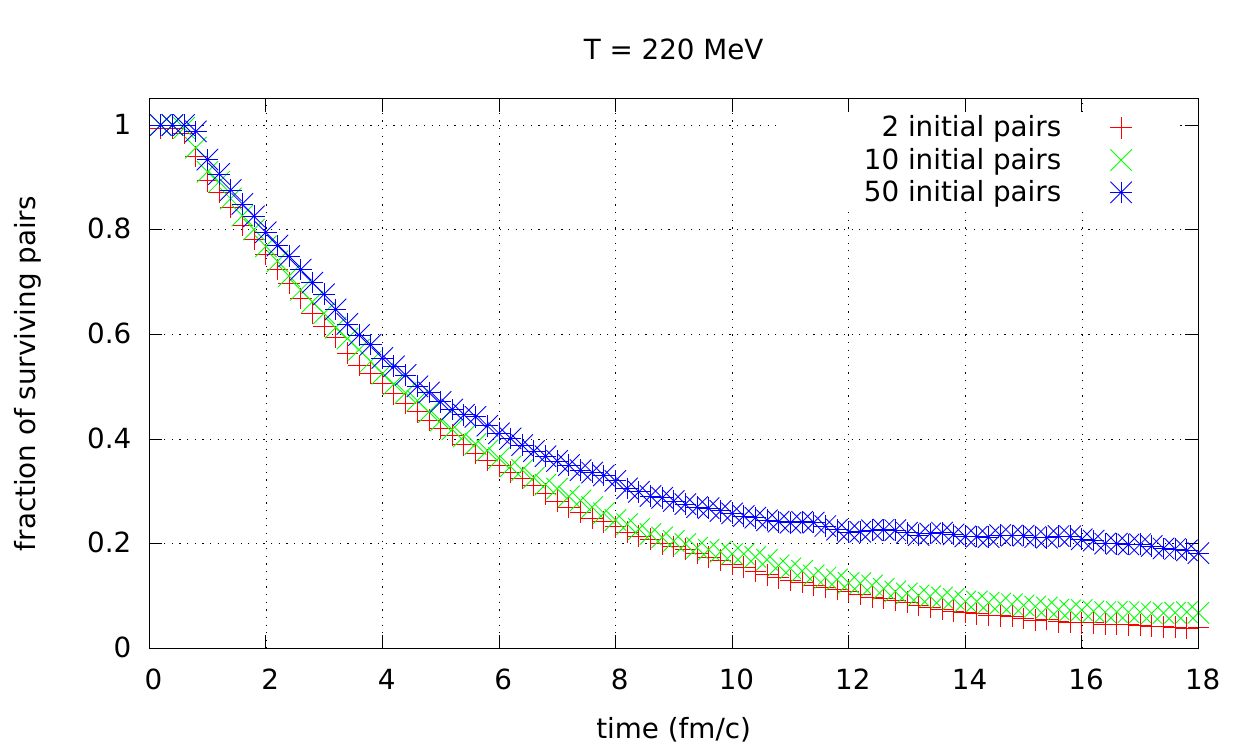}
\includegraphics[width=7.5cm]{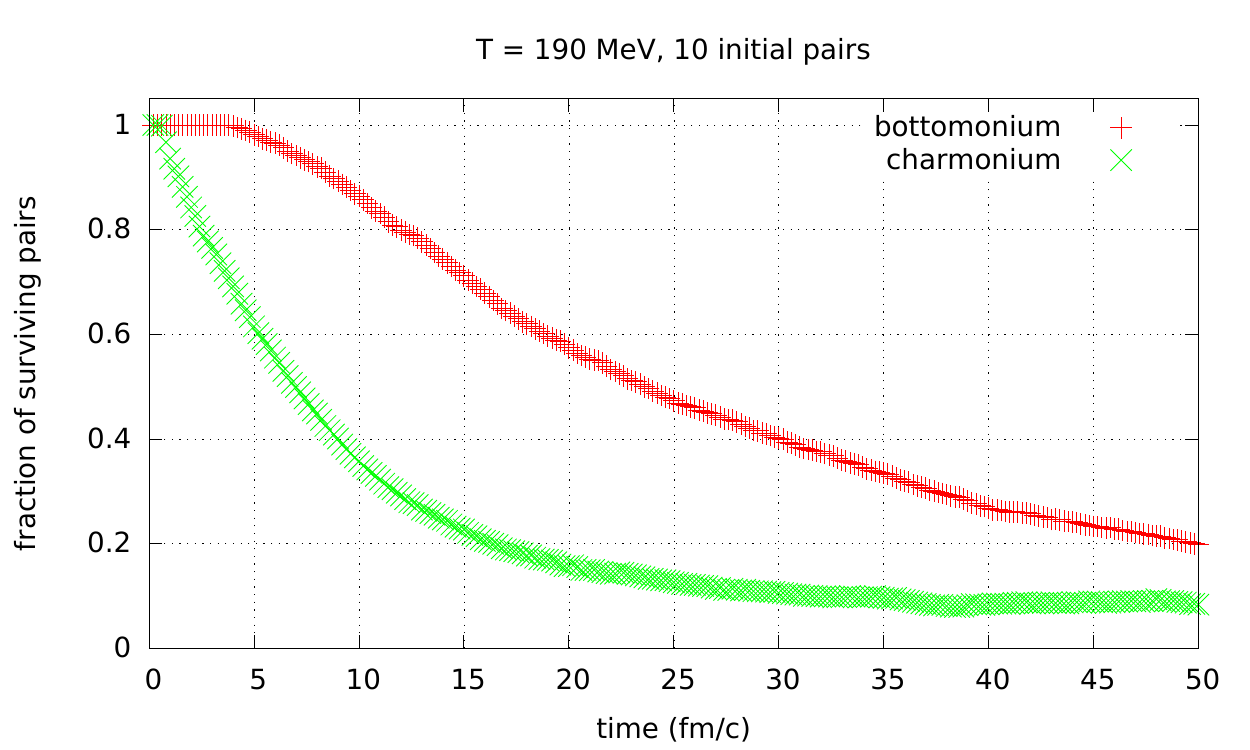}
\caption{Fraction of surviving pairs as a function of time for three different temperatures. The $N$ pairs ($N=2,10,50$) are prepared at $t=0$ as bound states as indicate earlier in the text. The short time behavior is dominated by dissociation. The process starts however only after some small delay $t_0\lesssim 1$ fm/c. This delay is much longer the the bottomonium, as revealed by the comparison displayed in the bottom-right panel: fraction of surviving $J/\Psi$ and $\Upsilon$,   for a system of $N=10$ $c$-$\bar c$ or $b$-$\bar b$ pairs. Simulations were performed in a periodic cubic box of side $4$ fm.}
\label{fig:surv}
\end{center}
\end{figure}

One can nevertheless push the discussion a bit further and quantify the process  in a simple way. Note first that  the recombination time is expected to go up when the number of particles decreases. This is indeed what we obtain from our simulations. We find that the average recombination time is, to a good approximation,  inversely proportional to the number of pairs present in the system: $t_{\rm rec}N\approx \lambda_R^{-1}\,$, with $\lambda_R$ a (temperature-dependent) recombination rate.
This effect is also (qualitatively) visible in Fig.~\ref{fig:surv} that displays the fraction of surviving $J/\Psi$ (and $\Upsilon$) particles as a function of time, for different number of pairs in the system: one notices that recombination events are more frequent in a system with a greater number of $c\bar c$ pairs.
 One may also observe that the effect of recombination becomes relatively more important as the temperature grows. This  is visible for instance from the development of a plateau suggestive of equilibrium that is most clearly seen at the highest temperature ($T=220$ MeV). Finally, the last panel of Fig.~\ref{fig:surv} compares the behaviors of $c\bar c$ and $b\bar b$ at a given temperature over a long time scale. One sees that there is a lapse of time before the $b\bar b$ bound state starts to ``feel'' the action of the thermal medium. This time delay $t_0$ is about $t_0\approx 4$ fm/c. A similar effect also occurs for charmonium, but for a smaller $t_0\lesssim 1$ fm/c. This dependence on the mass is a clear indication of the important role of the collisions in the dissociation process. 
 
 At the same time, the effect of the binding forces is certainly also present. This  we see indirectly by studying the cutoff dependence of the results. To that aim, we have repeated simulations for various values of the cutoff. As we have seen earlier, the dominant effect of a change in the cutoff is to change the depth of the potential. A larger cutoff leads to a deeper potential, and a longer lifetime, and this effect persists up to values of the order $\Lambda\approx 6$ GeV, above which it attenuates considerably.  In turns, this alters the recombination rates  since  the pairs with too short lifetimes are eliminated by the  procedure with which we identify bound pairs.

One may understand quantitatively the behaviors identified in Fig.~\ref{fig:surv} from a simple rate equation. Let us denote by $\lambda_D(T)$   the dissociation rate and by $\lambda_R(T)$ the recombination rate. Both are  functions of the temperature. 
The rate equation describing the time evolution of the number of surviving $Q\bar{Q}$ pairs $N(t)$ is (see also \cite{Thews:2005vj})
\be\label{eq:rate}
\deriv{N(t)}{t} = -\lambda_D N(t) + \lambda_R N_q(t)N_{\bar{q}}(t)\:,
\ee
where   $N_q=N_{\bar{q}}\,$ is the number of free heavy quarks (or antiquarks) in the plasma. Equation (\ref{eq:rate}) together with the initial condition $N(t=t_0)=N_0\,$, $N_{\bar q}(t_0)=N_q(t_0)=0$ can be analytically solved  for the fraction of surviving pairs:
\be\label{sol}
\frac{N(t)}{N_0} = \frac{1-\frac{\lambda_D}{\Omega}\tanh\left(\frac{\Omega}{2}(t-t_0)\right)}{1+\frac{\lambda_D}{\Omega}\tanh\left(\frac{\Omega}{2}(t-t_0)\right)}\:,\qquad t\geq t_0\:,
\ee
where  $\Omega\equiv\sqrt{\lambda_D(\lambda_D + 4\lambda_R N_0)}$ and we used  $N_q(t) = N_{\bar q}(t) = N_0-N(t)$. The time $t_0$ is the time at which dissociation starts, as defined earlier, and this time is chosen as the initial time when solving the rate equation (\ref{eq:rate}). 
\begin{table}[htdp]
\begin{center}
\begin{tabular}{ccc}
\hline
$T$ (MeV)  & $\lambda_D^{-1}\,(\mbox{fm/c})$  & $\lambda_R^{-1}\,(\mbox{fm/c})$ \\
\hline
 160       &    23     &  625 \\
 190       &   9.2  &  1000\\
 220       &   4.6      &  1350\\
 \hline
\end{tabular}
\caption{The inverse of the dissociation rate $\lambda_D^{-1}$ and the recombination rate $\lambda_R^{-1}$, for various temperatures.}
\end{center}
\label{tab:decay}
\end{table}

\begin{figure}
\begin{center}
\includegraphics[width=7.5cm]{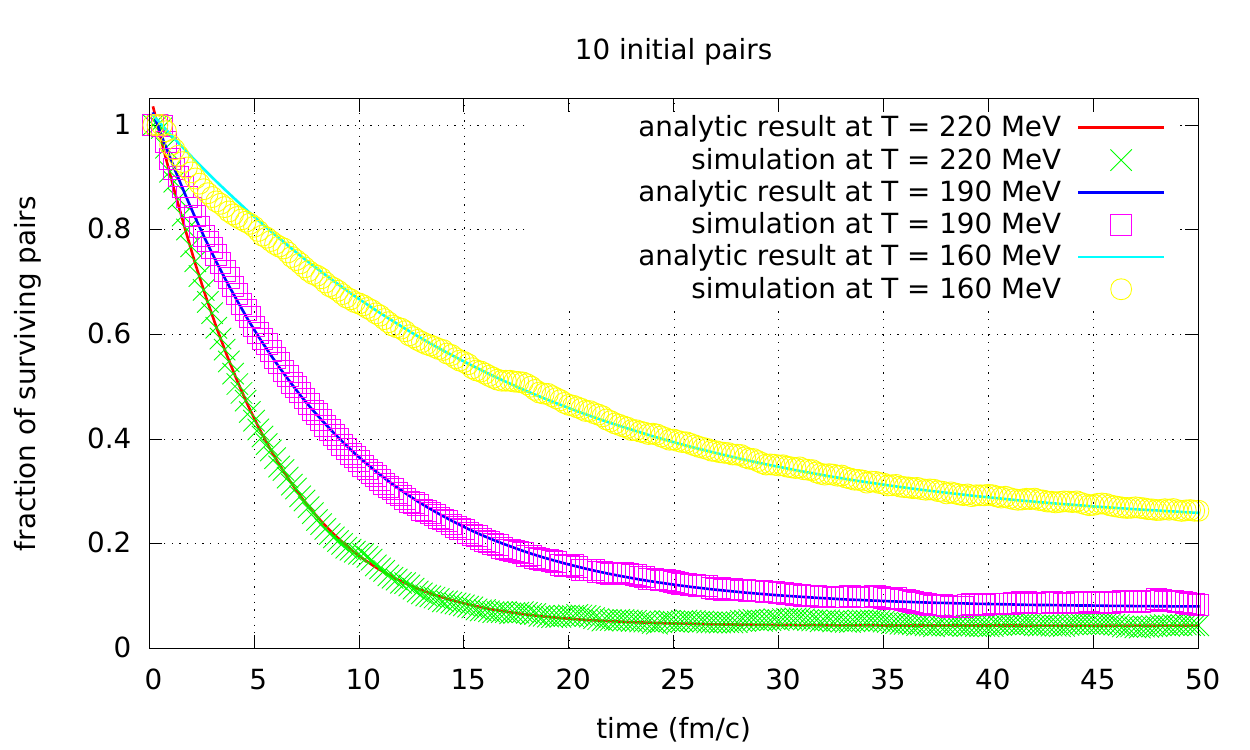}
\includegraphics[width=7.5cm]{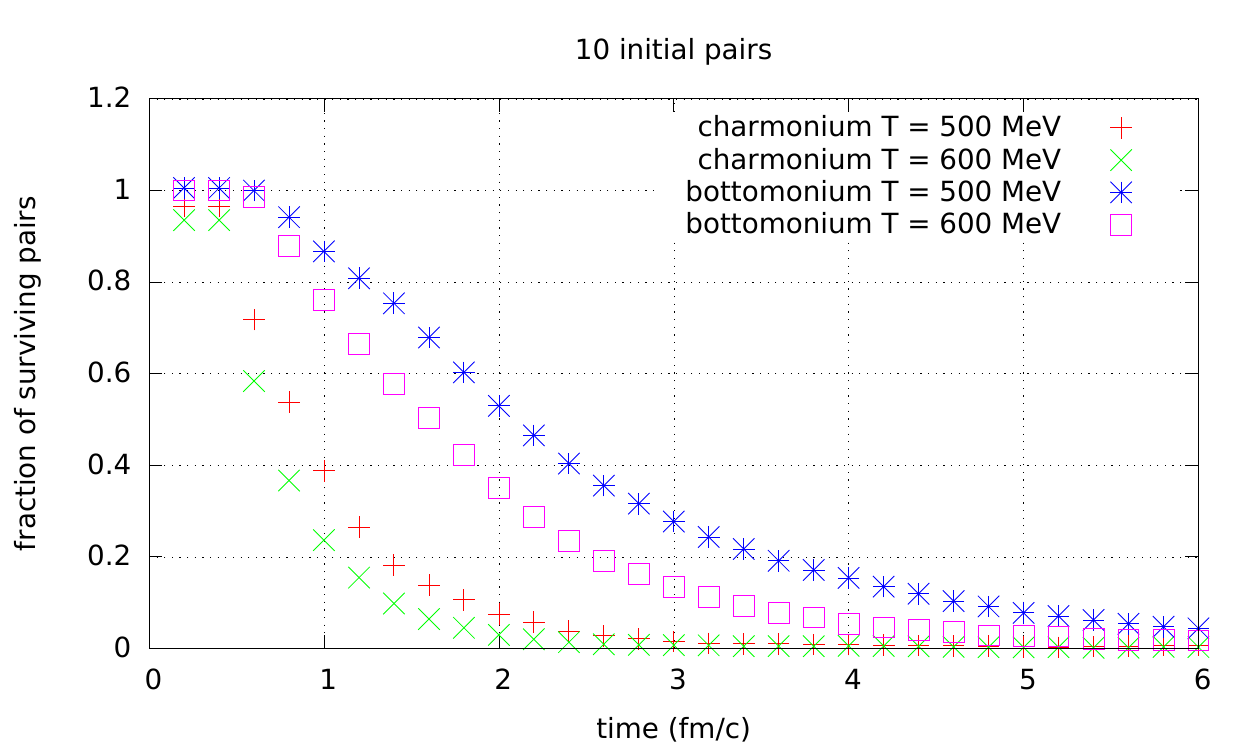}
\caption{Left panel: fit of the solution (\ref{sol}) of the rate equation for a system of $10$ initial $J/\Psi$ at three different temperatures. We notice that eq.(\ref{sol}) fits the curves very well with the recombination rates $\lambda_R$ equal to the ones derived in Fig.~\ref{fig:rec}. Right panel: high temperature behaviour of charmonium and bottomonium.}
\label{fig:fit}
\end{center}
\end{figure}

From the results of the simulations, and using  Eq.~(\ref{sol}), we can extract values for the dissociation and regeneration rates. The results are reported in Table~\ref{tab:decay} for a few temperatures. The quality of the fit can be seen on Fig.~\ref{fig:fit}. The values of the dissociation rate thus determined can be used to infer the lifetimes of the bound states. The values of the $J/\Psi$ lifetimes obtained from the present fit  agree with those previously obtained by analyzing the time evolution of the $J/\Psi$ radius (see Table~\ref{tab:lifetime}). The  analysis of the values of the dissociation rate just obtained suggests  that at a temperature $T=600$ MeV, the lifetime of the $J/\psi$ is still non vanishing, and is about .5 fm/c. At the same temperature, the lifetime of the $\Upsilon$ is about $1.5$ fm/c. The fit of bottomonium data gives essentially zero recombination rates already at $T=160$ MeV ($\,\lambda_R\sim (1\pm 1)\cdot 10^{-5}$ fm/c) and much smaller values of the dissociation rate as compared to the charmonium case. These numbers reflect of course  the greater stability of the $\Upsilon$ as compared to  the $J/\Psi$.

\section{Conclusions}

We have presented an approach that treats  in a unique framework most of the important aspects of the evolution of a collection of heavy quark-antiquark pairs propagating through a quark gluon plasma. The approach starts from first principles, and leads, through well defined approximations, to a complete dynamical description  with  a unified perspective on many different physical effects, usually treated with different models. Of course, several approximations are needed to arrive at tractable calculations. However these approximations can be improved, and their presence should not obscure  the overall consistency of the general scheme. We find it particularly important,  for instance, in view of their relevance  for the interpretation of the data,  to have the processes of dissociation and recombination treated on the same footing. 

 The main question that is addressed in this paper is of a general nature, it concerns the fate of a collection of heavy quark pairs in a hot and dense environment, with the possibility for these heavy particles to form bound states. This does not involve QCD dynamics in an essential way, and this is the main reason why we have restricted ourselves in this paper to Abelian plasmas where the same question can be addressed in a much simpler setting. Specific features of  QCD can be  implemented within the present scheme, with perhaps some approximations becoming less accurate. In particular, one  of the main approximations can be understood as a weak coupling approximation, which consists in neglecting the non linear coupling of the gauge (Coulomb) field with which the particles interact. The field fluctuations are then Gaussian, which allows for a simple calculation of  the influence functional in terms of a 2-point function that characterizes entirely the effect of the plasma on the heavy particles.  In QCD, the non linear couplings are not as strongly suppressed as in QED, and the approximation may be less accurate.

Further approximations lead to a classical treatment of the dynamics in terms of a Langevin equation, in which the noise term accounts for the effect of the collisions between the heavy particles and the plasma constituents. This noise terms depends on the positions of the heavy quarks at each time steps. This dependence is an important aspect of the dynamics. 

The simulations presented in this paper are the first in a program that  can be improved in many ways. Some of the approximations that have been made can easily be relaxed, such as for instance the Abelian approximation, as we have discussed already. The  classical treatment of the dynamics through a Langevin equation could be improved, e.g. including the leading order quantum corrections to the Langevin equation, as shown in \cite{beccara}.  The basic  ingredients such as the transport coefficients, can be calculated with greater accuracy. Finally, once some of these improvements are implemented, more realistic phenomenological applications can be envisaged.  We hope to be able to report on some of these developments soon.

\vspace{1cm}

\noindent{\bf Acknowledgements}

  This research is supported by the European Research Council under the Advanced Investigator Grant ERC-AD-267258. JPB acknowledges early discussions on some aspects of this project with F.~Dominguez and B. Wu, as well as with Y.~Akamatsu.  GG acknowledges support by Istituto Nazionale di Fisica Nucleare through the ``Supercalcolo'' agreement with Fondazione Bruno
Kessler. Computer simulations were performed on the KORE computing cluster at FBK.

\appendix

\section{Taylor expansion of the Influence Functional}\label{expand}

In this Appendix, we perform the Taylor expansion of $\Phi[\Q] = \Phi_{_{QQ}}[\Q]+\Phi_{_{\bar Q\bar Q}}[\Q]+\Phi_{_{Q \bar Q}}[\Q]$ obtained from Eqs.~(\ref{QQ}) and (\ref{QbarQ}) of section \ref{low} after performing the change of coordinates of Eq.~(\ref{substitution}).
We first analyze the contribution $ \Phi_{_{QQ}}[\Q]$ involving only the heavy quarks. We have
\be\label{toexp}
&&\Phi_{_{QQ}}[\Rg,\Yg] = \frac{g^2}{2} \sum_{i,j=1}^N \int_{t_i}^{t_f} \diff{t} 
\left\{ V\left(\rr_j-\rr_i-\frac{\y_j}{2}+\frac{\y_i}{2}\right)-V\left(\rr_j-\rr_i+\frac{\y_j}{2}-\frac{\y_i}{2}\right)\right.\nn\\
&&\left.-\iu W\left(\rr_j-\rr_i-\frac{\y_j}{2}+\frac{\y_i}{2}\right)-\iu W\left(\rr_j-\rr_i+\frac{\y_j}{2}-\frac{\y_i}{2}\right) + 2\,\iu W\left(\rr_j-\rr_i+\frac{\y_j}{2}+\frac{\y_i}{2} \right)\right.\nn\\
&&\left.-\frac{\beta}{2}\left(\dot{\rr}_j+\dot{\rr}_i+\frac{\dot{\y}_j}{2}-\frac{\dot{\y}_i}{2}\right)\cdot \left(\frac{1}{2}\frac{\partial}{\partial \rr_j}+\pderiv{}{\y_j}\right) W\left(\rr_j-\rr_i+\frac{1}{2}\y_j+\frac{1}{2}\y_i\right)
\right\},\nn\\
\ee
where the time dependence is hidden in the coordinates $\Rg$ and $\Yg$. The dot symbol in this expression, as well as in the rest of this section, denotes a scalar product and involves the three cartesian components of the vectors. 
We want to expand the expression (\ref{toexp}) to second order in $\y$. To do so, we use the well-known Taylor expansion of a scalar function $f$ of a $n$-dimensional vector $\x$, 
\be\nn
f(\x)=f(\ag)+(\x-\ag)\cdot\nab f(\ag)+\frac{1}{2}~(\x-\ag)\cdot \h^f(\ag)\cdot(\x-\ag)+\dots
\ee
where $\x=(x_1,\dots,x_n)$, and the gradient and Hessian matrix are given, as usual,  by 
\beq\label{Hessiandef}
 \nabla_\alpha f(\ag):=\left.\pderiv{f(\x)}{x_\alpha}\right|_{\x=\ag},\qquad \h_{\alpha\beta}^f(\ag):=\left.\frac{\partial^2 f(\x)}{\partial x_\alpha\partial x_\beta}\right|_{\x=\ag}\;,\qquad \alpha,\beta=1,\cdots,n.
 \eeq
  By applying this formula to the first line of Eq.~(\ref{toexp}) we get (with $\rr_{ji}\equiv \rr_j-\rr_i$, $\y_{ji}\equiv \y_j-\y_i$)
  \beq
V(\rr_{ji}-\frac{1}{2} \y_{ji})-V(\rr_{ji}+\frac{1}{2} \y_{ji})=-\y_{ji}\cdot\nab V(\rr_{ji}).
\eeq
  Similarly\footnote{The Hessian matrix of $W$ is the only such matrix in the present discussion, so we denote it simply by $\h$, without any explicit reference to $W$ in the notation. That is, in the notation of Eq.~(\ref{Hessiandef}), $\h\equiv\h^{_{W}}$.},
  \beq
 -W(\rr_{ji}-\frac{1}{2} \y_{ji})-W(\rr_{ji}+\frac{1}{2} \y_{ji}) =-2W(\rr_{ji})-\frac{1}{4}\,\y_{ji}\cdot\h(\rr_{ji})\cdot \y_{ji},
 \eeq
 and (with $\tilde \y_{ji}\equiv \y_j+\y_i$)
 \beq
 W\left(\rr_{ji}+\frac{\tilde\y_{ji}}{2}\right)=W(\rr_{ji})+\frac{\tilde\y_{ji}}{2}\cdot\nab W(\rr_{ji})+\frac{1}{8}\,\tilde\y_{ji}\cdot\h\cdot \tilde\y_{ji}.
 \eeq
Note that the middle term in the right hand side of the last equation will disappear in the summation over $i$ and $j$, since it is antisymmetric ($\nab W(\rr_{ji})=-\nab W(\rr_{ij})$)\footnote{When using regularized potentials $V(\rr)$ ad $W(\rr)$ such that $\nab V(\rr) $ and $\nab W(\rr)$ both vanish at $\r=0$, the same cancellation holds for the terms with $i=j$.}.
  
Let us now consider the terms that involve the time derivative. We write this as 
\beq
-\frac{\beta}{2}\left(\dot{\tilde{\rr}}_{ji}+\frac{\dot{\y}_{ji}}{2}\right)\cdot \frac{\partial}{\partial \rr_{ji}} W\left(\rr_{ji}+\frac{1}{2}\tilde\y_{ji}\right)
\eeq
and use the expansion of $W$ above. When keeping only the symmetric terms, i.e., those which survive in the summation over $i$ and $j$, this yields
\beq
-\frac{\beta}{4}\left(    \dot{\tilde{\rr}}_{ji}\cdot \h\cdot  \tilde\y_{ji} +{\dot{\y}_{ji}} \cdot\nab W\left(\rr_{ji} \right) \right).
\eeq
At this point,  we note that one can write  $\dot{\y}_{ji}\cdot \nab W\left(\rr_{ji}\right)$ as $-\y_{ji}\cdot \h(\rr_{ji})\cdot \dot\rr_{ji}$ after integrating by part in the integral over time appearing in $\Phi_{_{QQ}}\,$. The boundary terms coming from this integration by parts vanish because the coordinates $\Q_1$ and $\Q_2$ coincide at both  ends  of the Schwinger-Keldysh contour, that is
\be
\y_j(t_f)=\q_{j,1}(t_f)-\q_{j,2}(t_f)=0, \qquad \y_j(t_i)=\q_{j,1}(t_i)-\q_{j,2}(t_i)=0 .\nn
\ee
Collecting all intermediate results, we get
\beq
\Phi_{_{QQ}}[\Rg,\Yg] &=& \frac{g^2}{8} \sum_{i,j=1}^N \int_{t_i}^{t_f} \diff{t}\left[-4\,\y_{ij}\cdot \nab V(\rr_{ij}) +\iu\left(\tilde\y_{ji}\cdot\h\cdot \tilde\y_{ji}-\y_{ji}\cdot\h\cdot \y_{ji}\right) +\right.\nn\\
&&\left.+\,\beta\left(\y_{ji}\cdot \h(\rr_{ij})\cdot \dot\rr_{ji} - \tilde\y_{ji}\cdot \h(\rr_{ji})\cdot \dot{\tilde{\rr}}_{ji} \right)\right], 
\eeq
which we can rewrite as  
\be
\Phi_{_{QQ}}[\Rg,\Yg] = -\frac{g^2}{2}\!\!\sum_{i,j=1}^N \!\int_{t_i}^{t_f} \!\!\diff{t}\!\left[2\,\y_i\cdot \nab V(\rr_{ij}) -\iu\,\y_i\cdot \h(\rr_{ij})\cdot \y_j + \beta\,\y_i\cdot \h(\rr_{ij})\cdot \dot{\rr}_j \right].\nn\\
\ee
The result for $\Phi_{_{\bar Q\bar Q}}$ is obtained trivially from $\Phi_{_{QQ}}$ via the  change of variables $\y\rightarrow \bar\y, \rr\rightarrow \bar\rr\,$. Let us then consider the expansion of the remaining term, $\Phi_{_{Q\bar Q}}[\Rg,\Yg]$. We have
\be\label{toexp2b}
&&\Phi_{_{Q\overline Q}}[\Rg,\Yg] = -g^2\sum_{i,j=1}^N \int_{t_i}^{t_f} \diff{t} 
\left\{ V\left(\rr_{ji}-\frac{1}{2}\y_{ji}\right)-V\left(\rr_{ji}+\frac{1}{2}\y_{ji}\right)\right.\nn\\
&&\left.-\iu W\left(\rr_{ji}-\frac{1}{2}\y_{ji}\right)-\iu W\left(\rr_{ji}+\frac{1}{2}\y_{ji}\right)+ \iu W\left(\rr_{ji}+\frac{1}{2}\tilde\y_{ji} \right)+ \iu W\left(\rr_{ji}-\frac{1}{2}\tilde\y_{ji} \right)\right.\nn\\
&&\left.-\frac{\beta}{2}\left(  \dot{\bar{\rr}}_i-\frac{\dot{\bar\y}_i}{2}\right)\cdot \frac{\partial}{\partial \rr_{ji}}   W\left(\rr_{ji}+\frac{1}{2}\tilde\y_{ji}\right)+\frac{\beta}{2}\left(\dot{\bar{\rr}}_i+\frac{\dot{\bar\y}_i}{2}\right)\cdot \frac{\partial}{\partial \rr_{ji}}  W\left(\rr_{ji}-\frac{1}{2}\tilde\y_{ji})\right)
\right\},\nn\\
\ee
with now $\rr_{ji}\equiv\rr_j-\bar\rr_i$, $\y_{ji}\equiv\y_j-\bar\y_i$ and $\tilde\y_{ji}\equiv\y_j+\bar\y_i$.
By using similar manipulations as above, 
 one finds that the last two terms contribute
\be
\frac{\beta}{2}\left(  \dot{\overline\y}_i\cdot\nab W(\rr_{ji}) - \dot{\bar\rr}_i\cdot\h(\rr_{ji})\cdot \tilde\y_{ji}  \right)=
-\frac{\beta}{2}\left( \bar\y_i\cdot \h(\rr_{ji})\cdot \dot\rr_{ji} + \dot{\bar\rr}_i\cdot\h(\rr_{ji})\cdot\tilde\y_{ji}  \right),\nn\\
\ee
where we have used an integration by part. 
Moving up to the second line, we get
\beq
\iu\,W\left(\rr_{ji}+\frac{1}{2}\tilde\y_{ji}) \right)+ \iu\,W\left(\rr_{ji}-\frac{1}{2}\tilde\y_{ji}) ) \right)=2\iu W(\rr_{ji})+\frac{\iu}{4} \tilde \y_{ji}\cdot\h(\rr_{ji})\cdot\tilde\y_{ji}
\eeq
and  
\beq
-\iu\,W\left(\rr_{ji}+\frac{1}{2}\y_{ji}) \right)- \iu\,W\left(\rr_{ji}-\frac{1}{2}\y_{ji}) ) \right)=-2\iu\, W(\rr_{ji})-\frac{\iu}{4}  \y_{ji}\cdot\h(\rr_{ji})\cdot\y_{ji}
\eeq
As for the first line, it yields simply
\beq V\left(\rr_{ji}-\frac{1}{2}\y_{ji})\right)-V\left(\rr_{ji}+\frac{1}{2}\y_{ji})\right)=-\y_{ji}\cdot\nab V(\rr_{ji}).
\eeq
Collecting all the intermediate results, we can then rewrite the influence functional as follows
\be\label{toexp4}
\Phi_{_{Q\overline Q}}[\Rg,\Yg] &=& - {g^2} \sum_{i,j=1}^N \int_{t_i}^{t_f} \diff{t} 
\left\{ -\,(\y_{j}-\bar\y_i)\cdot\nab V(\rr_{j}-\bar\rr_i)+\iu\, \y_{j}\cdot\h(\rr_j-\bar\rr_i)\cdot\bar\y_{i}\right.\nn\\
&&\left.-\frac{\beta}{2}\left({\bar\y}_i\cdot\h(\bar\rr_i-\rr_j)\cdot \dot\r_j  +\y_j\cdot\h(\rr_j-\bar\rr_i)\cdot \dot{\bar\rr}_i\right)
\right\}.\nn\\
\ee
By collecting the Taylor expansions of $\Phi_{_{QQ}}, \Phi_{_{\bar Q\bar Q}}$ and $\Phi_{_{Q\bar Q}}$ derived in this appendix, and using the definitions (\ref{Force}) and (\ref{hessian}),  one easily obtains the equations~(\ref{langpath}, \ref{L}) of the main text

\section{Derivation of the generalized Langevin equation}\label{appendixlangevin}

In this Appendix, we show that the dynamics encoded in the path integral (\ref{langpath})  is equivalent to that described by the generalized Langevin equation (\ref{finallangev}). 
Let us start by considering the Langevin equation for a particle of mass $M$  moving in an  $N$-dimensional space: 
\be\label{L2}
M\,\ddot{r}_i = -M\gamma_{ij}\,\dot{r}_j + f_i(\rr) + \xi_i(t)\:,\qquad i=1,\dots,N, 
\ee
where $r_i$ denotes a coordinate of the particle, $\dot r_i$ and $\ddot r_i$ its first and second time derivatives,
$f_i$ is an external deterministic force, and $\xi_i$ a white stochastic force with the following properties
\be
\langle~\xi_i(t)~\rangle_{\xi}=0,\qquad
\langle~\xi_i(t)~\xi_j(t')~\rangle_{\xi} = \lambda_{ij}\,\delta(t-t^\prime),\qquad \lambda_{ij}=2MT\gamma_{ij}.
\ee
Here $\gamma$  is a real symmetric matrix, and   we have used Einstein's relation  between the noise and the dissipative terms. 
The equation that we need to consider is a generalization of Eq.~(\ref{L2}) in which  the matrix $\gamma_{ij}$ (and hence $\lambda_{ij}$) depends on the position $\rr$ of the heavy particle. It is of the form
\be\label{Lg}
M\,\ddot{r}_i = -M\gamma_{ij}(\rr)\,\dot{r}_j + f_i(\rr) + \xi_i(\rr,t)\,.
\ee
with a  so-called multiplicative noise 
\beq
\xi_i(\rr,t):=w_{ij}(\rr)~\xi_j(t), \qquad  w_{ik}(\rr)w_{jk}(\rr)=\lambda_{ij}(\rr), \qquad 
\langle~\xi_i(t)~\xi_j(t')~\rangle_{\xi} = \delta_{ij}\,\delta(t-t^\prime).\nn\\
\eeq
An equation such as Eq.~(\ref{Lg}) may suffer from discretization ambiguities in the case where the inertia term, the left hand side of the equation, is ignored, leading to the so-called ``overdamped'' Langevin equation (see for instance \cite{LauLub}). These ambiguities reside in the choice of the point $\rr$ where the noise is evaluated when one solves the stochastic equation. 
One may indeed choose to evaluate the noise $w(\rr)$ at any point  $\rr$ between $\rr(t)$ and $\rr(t+\Delta t)$, where $\Delta t$ is the discrete time step. This leads to an uncertainty of order 
\be\label{error}
\frac{\rmd w}{\rmd \rr}\cdot\dot \rr~\Delta t~\xi, 
\ee 
with $w(\rr)$ assumed to be a smooth function of $\rr$. 
In the overdamped case, we have $\dot \rr\sim \xi\sim 1/\sqrt{\Delta t}$ so that the uncertainty is of order unity and remains finite as $\Delta t\to 0$. 
 However, as discussed in \cite{Arnold:1999uza}, such ambiguities may not appear when the inertial term is present, which is the case of interest in the present discussion. This is because, one can rewrite the equation (\ref{Lg}) as a set of two coupled equations, 
\be\label{division}
&& \dot{\rr} = \v \nn\\
&& M\dot{\v} = -M\bmgamma(\rr)\cdot \v + {\bf f}(\rr) + \bmw(\rr)\cdot\bmxi(t)\,.
\ee
In this case, while $\dot v \sim \xi\sim 1/\sqrt{\Delta t}$, $v$ itself remains finite, and so does $\dot r$. It follows that the uncertainty (\ref{error}) is now of order $\sqrt{\Delta t}$ and it vanishes as $\Delta t\to 0$.\\

We shall then proceed to the discretization of the system of equations (\ref{division}), and to be specific, we shall use the Ito convention, where the noise is estimated at the position of the particle before the time step considered. The discretized form of Eqs.~(\ref{division}) reads then 
\be\label{langdiscr}
&&\rr^{(n)} - \rr^{(n-1)} = \Delta t~\v^{(n-1)} + \Delta t \int_{t_{n-1}}^{t_n}\diff{s}~\bmzeta(s)\nn\\
&&M\left(\v^{(n)} - \v^{(n-1)}\right) = -M\Delta t\,\bmgamma^{(n-1)}\cdot\v^{(n-1)} + \Delta t \,{\bf f}^{(n-1)} + \int_{t_{n-1}}^{t_n}\diff{s}~\bmxi^{(n-1)}(s)\,,\nn\\
\ee
with initial conditions $\rr^{(0)}=\rr_0$ and $\v^{(0)}=\v_0$. In the equations above, we have set $\bmgamma^{(n-1)}=\bmgamma(\rr^{(n-1)})$, ${\bf f}^{(n-1)}={\bf f}(\rr^{(n-1)})$ and $\bmxi ^{(n-1)}(s)=\bmxi(\rr^{(n-1)},s)$ in order to simplify the notation. We have 
\be\label{devstand}
\langle~\xi_i^{(n)}(t)~\xi_j^{(n)}(t')~\rangle = \gamma_{ij}^{(n)}\,\delta(t-t').
\ee
To write Eqs.~(\ref{langdiscr}), we  have divided the time interval $[0,t]$ into $\bar n$ time step $\Delta t$, $\Delta t := t_n-t_{n-1}\,$, $n=1,\dots,\bar{n}$. We have also  introduced in the first equation an auxiliary white Gaussian noise $\zeta_i(t)$ with properties (see also \cite{Hanggi} for a similar procedure) 
\be
\langle~\zeta_i(t)~\rangle=0,\qquad
\langle~\zeta_i(t)~\zeta_j(t')~\rangle = \mu~\delta_{ij}~\delta(t-t')\,.
\ee
The additional factor $\Delta t$ in front of the  integral of the noise in the first equation (\ref{langdiscr})  ensures that $\dot{\rr}=\v$ in the limit $\Delta t\rightarrow 0$, thus avoiding any discretization ambiguity.\footnote{In \cite{Hanggi} the extra $\Delta t$ is not used, but $\mu$ is eventually sent to zero. The advantage of keeping the factor $\Delta t$ explicit is that it makes obvious that all potential discretization ambiguities disappear as $\Delta t\to 0$.} 

The conditional probability for the particle to be found in the configuration $\left(\rr^{(n)}, \v^{(n)}\right)$ at time $t_n$, given that it is in the  configuration $\left(\rr^{(n-1)}, \v^{(n-1)}\right)$ at time $t_{n-1}$, can be written as \cite{LauLub}
\be\label{Pelem}
P\left(\rr^{(n)}, \v^{(n)}, t_n\,|\,\r^{(n-1)}, \v^{(n-1)}, t_{n-1}\right)=\langle\,\delta\left(\rr^{(n)}-\rr_{sol}^{(n)}\right)\rangle_{\zeta}~\langle\,\delta\left(\v^{(n)}-\v_{sol}^{(n)}\right)\,\rangle_{\xi} ,\nn\\
\ee
where $\rr_{sol}^{(n)}=\r_{sol}(t_n; \rr^{(n-1)}, t_{n-1})$ and $\v_{sol}^{(n)}=\v_{sol}(t_n; \v^{(n-1)}, t_{n-1})$ are the solutions of the discretized equations (\ref{langdiscr}) for a given realization of the (independent) noises $\xi$ and $\zeta$, and given values of  $\v^{(n-1)}$ and $\rr^{(n-1)}$ at time $t_{n-1}$.
The probability of having a final configuration $(\rr^{(\bar{n})}, \vg^{(\bar{n})})=(\rr , \vg)$ at time $t_{\bar{n}}=t$ given an initial configuration $(\rr^{(0)}, \vg^{(0)})=(\rr_0,\v_0)$ at time $t_0=0$ is  given in terms of the probability  (\ref{Pelem}) of an elementary step by
\be\label{probafull}
&&\!\!P\left(\rr, \vg, t~|~\rr_0, \vg_0, t_0\right) =\nn\\
&&\!\!= \prod_{n=1}^{\bar{n}-1}\int\diff^{_{N}}{\rr}^{(n-1)}\int\diff^{_{N}}{\v}^{(n-1)}~P\left(\rr^{(n)}, \v^{(n)}, t_{n}~|~\rr^{(n-1)}, \v^{(n-1)}, t_{n-1}\right)\:.
\ee
This formula is the starting   point for building the path integral. In order to evaluate the average over the noises of the delta functions in Eq.~(\ref{Pelem}), it is convenient to  define the following $N$-dimensional vectors
\be
&&\!\!\!\!\!\!\!\!\!\!\bmg^{(n)}:= \rr^{(n)} -\rr^{(n-1)} - \Delta t~\v^{(n-1)} - \Delta t\int_{t_{n-1}}^{t_n}\!\!\!\!\!\!\!\diff{t}~\bmzeta(t)\\
&&\!\!\!\!\!\!\!\!\!\!\bmh^{(n)}:= M\,(\v^{(n)} -\v^{(n-1)}) + M \Delta t~\bmgamma^{(n-1)}\cdot\v^{(n-1)} - \Delta t~{\bf f}^{(n-1)} - \int_{t_{n-1}}^{t_n}\!\!\!\!\!\!\!\diff{t}~\bmxi^{(n-1)}(t).\nn
\ee
These functions vanish respectively when $\rr^{(n)}= \rr_{sol}^{(n)}$ and $\v^{(n)}= \v_{sol}^{(n)}$, and we have 
\be\label{deltafunctions}
\delta\left(\rr^{(n)}-\rr_{sol}^{(n)}\right) =\delta\left(\bmg^{(n)}\right),\qquad
\delta\left(\v^{(n)}-\v_{sol}^{(n)}\right)= M^N\delta\left(\bmh^{(n)}\right).
\ee
In order to calculate the noise averages, we use the Fourier representation of these delta functions. We get then 
Taking the mean value of these quantities, we get
\be
\langle\delta\left(g_k^{(n)}\right)\rangle_{\zeta} &= &\int_{-\infty}^{+\infty}\frac{\diff{z^{(n)}_k}}{2\pi}~\eu^{-\iu z^{(n)}_k (r^{(n)}_k -r^{(n-1)}_k - \Delta t~v^{(n-1)}_k)}\langle\eu^{\,\iu z^{(n)}_k\Delta t\,\int_{t_{n-1}}^{t_n}\diff{s}~\zeta_k(s)}\rangle_{\zeta}\,,\nn\\
&= &\int_{-\infty}^{+\infty}\frac{\diff{z^{(n)}_k}}{2\pi}~\eu^{-\iu z^{(n)}_k (r^{(n)}_k -r^{(n-1)}_k - \Delta t~v^{(n-1)}_k)},
\ee
where we have exploited the presence of the explicit factor $\Delta t$ in order to evaluate the average of the last exponential factor to linear order in $\Delta t$, where it reduces to unity. Similarly,   
\be
\langle\delta\left(h_k^{(n)}\right)\rangle_{\xi} &=& \int_{-\infty}^{+\infty}\frac{\diff{y^{(n)}_k}}{2\pi}~\eu^{-\iu\,y^{(n)}_k (M(v^{(n)}_k -v^{(n-1)}_k) + M\Delta t \,\gamma_{kj}^{(n-1)}\,v^{(n-1)}_j - \Delta t~f_k^{(n-1)})}\nn\\
&\times&\langle\,\eu^{\,\iu y^{(n)}_k\,\int_{t_{n-1}}^{t_n}\diff{s}~\xi_k^{(n-1)}(s)}\rangle_{\xi},
\ee
where now we need to push the expansion of the last exponential to second order
\be
&&\langle\,1-\frac{1}{2}\,y^{(n)}_ky^{(n)}_j\int_{t_{n-1}}^{t_n}\diff{s}\int_{t_{n-1}}^{t_n}\diff{u}\,\xi_k^{(n-1)}(s)\,\xi_j^{(n-1)}(u)\,\rangle_{\xi} \nn\\
&&=  1-\frac{1}{2}~y^{(n)}_ky^{(n)}_j\,\lambda_{kj}^{(n-1)}~\Delta t \approx \eu^{-\frac{1}{2}~y^{(n)}_ky^{(n)}_j\,\lambda_{kj}^{(n-1)}~\Delta t}.
\ee
The probability $P(\rr, \vg, t~|~\rr_0, \vg_0, t_0)$ of Eq.~(\ref{probafull}) can therefore be written as (to within the factor $M^N$ coming from the delta functions (\ref{deltafunctions}) and that can be absorbed in the normalization)
\be
&&\!\!\!\!\!\!\!\!\!\!\!\prod_{n=1}^{\bar{n}-1}\int\!\diff{\rr}^{(n-1)}\!\!\int\!\diff{\v}^{(n-1)}\!\!\int_{-\infty}^{+\infty}\!\frac{\diff{\y^{(n)}}}{2\pi}\int_{-\infty}^{+\infty}\!\frac{\diff{\z^{(n)}}}{2\pi}\exp\left[-\iu \Delta t\, \z^{(n)}\cdot \left(\frac{\rr^{(n)} -\rr^{(n-1)}}{\Delta t}-\v^{(n-1)}\right)\right]\nn\\
&&\!\!\!\!\!\!\!\!\!\!\!\times\exp\left[-\iu\,\Delta t\,\y^{(n)}\cdot \left(M\,\frac{\v^{(n)} -\v^{(n-1)}}{\Delta t} + M\bmgamma^{(n-1)}\cdot\v^{(n-1)} - {\bf f}^{(n-1)}\right)-\frac{1}{2}\,\y^{(n)}\cdot\bmlambda^{(n-1)}\cdot \y^{(n)}\right].\nn\\
\ee
At this point we note that we can integrate over $\z$, thereby reconstructing the delta function $\delta\left[\dot{\rr}(t) - \v(t)\right]$, which removes the integration over the velocity $\v$. Note that there is no integration over $\z_0$ nor $\z_{\bar n}$, in line with the fact that $\v_0$ and $\v_{\bar n}$ are fixed by the initial and final conditions on the paths. Note also that the relevant path are differentiable, with $\dot\r=\v$ finite.  After integrating over $\v_0$ and $\v$, and sending $\Delta t\to 0$,  the remaining path integrals over $\rr$ and $\y$ yield the probability $P(\rr_f, t_f~|~\rr_i, t_i)$ in the form 
\be\label{finalpath}
P(\rr_f, t_f~|~\rr_i, t_i) = \int\D\rr\int\D\y\,\exp\left[ \int_{t_i}^{t_f}\diff{t}~{\cal L}(\rr,\y) \right],
\ee
with
\be
{\cal L}(\rr,\y)=
 -\iu\,\y\cdot\left(M\ddot{\rr} + M\bmgamma(\rr)\cdot\dot{\rr} - \mathbf{f}(\rr)\right)-\frac{1}{2}\y\cdot\bmlambda(\rr)\cdot\y.
\ee
where we should remember that there is no integration on the end point of the $\y$ path integral, or equivalently that $\y(t_f)=\y(t_i)=0$.
The structure of this expression is identical to that of the conditional probability (\ref{langpath}) derived in section \ref{sec:genLang}.

\section{Numerical algorithm}\label{algorithm}

In order to perform numerical simulations we use an explicit second-order algorithm\footnote{By second-order algorithm we mean that the convergence of the algorithm is of the order of $\Delta t^2$, with $\Delta t$ the time step of the simulation.} which requires the evaluation of a single function at each  time step. This algorithm can be summarized as follows. At each time step:
\begin{itemize}
  \item use an orthogonal transformation to pass to coordinates for which the real and symmetric matrix $\bmgamma$ is diagonal;
  \item perform the stochastic Verlet algorithm (see below);
  \item come back to original coordinates.
\end{itemize}

Let us detail the second step. After diagonalizing $\bmgamma$, we are left with with $N$ independent stochastic equations of the Ornstein-Uhlenbeck type \cite{Ornstein1930}
\be\label{diagonallang}
\dot{p} = -{\gamma(r)}\,p + f(r) + \sqrt{2MT\gamma(r)}\,\xi(t)\:,
\ee
\be\label{diagonallang2}
\dot r=\frac{p}{m},
\ee
with 
\be
\langle\,\xi(t)\,\rangle = 0,\qquad 
\langle\,\xi(t)\,\xi(t')\,\rangle = \delta(t-t').
\ee
Eq.~(\ref{diagonallang}) can be written as
\be
\diff{p} = - {\gamma(r)}\,p\,\diff{t} + f(r)\diff{t} + \sqrt{2MT\,\gamma(r)}\,\diff{\xi(t)},
\ee
which admits the exact solution \cite{Ornstein1930}
\be\label{exactsolution}
p_{t+\Delta t} = p_t\,\eu^{-\alpha} + \frac{f}{\gamma}\left(1-\eu^{-\alpha}\right) + \sqrt{MT\left(1-\eu^{-2\,\alpha}\right)}\,\xi_t\, , 
\ee
where $\alpha(r)=\gamma(r){\Delta t}$. This result enables us to write the propagation of Eqs.~(\ref{diagonallang}) and (\ref{diagonallang2}) in a manner similar to the Verlet algorithm used in Newtonian dynamics \cite{Scherer}
\be
r_+ &=& r_t + \frac{p_t}{2\,M}\,\Delta t\nn\\
p_{t+\Delta t} &=& (1-\alpha(r_+))\,p_t + f(r_+)\Delta t + \sqrt{2MT\,\alpha(r_+)}\,\xi_t\nn\\
r_{t+\Delta t} &=& r_+ +\frac{p_{t+\Delta t}}{2M}\,\,\Delta t. 
\ee
which is often referred to as the  stochastic Verlet algorithm \cite{Tony}.

\end{document}